\documentclass[twoside]{elsart}
\usepackage{amsmath}

\usepackage{graphicx}

\usepackage{amssymb}

\begin{document}

\def\d{\partial}
\def\um{\,\mu{\rm  m}}
\def\mm{\,   {\rm mm}}
\def\cm{\,   {\rm cm}}
\def \m{\,   {\rm  m}}
\def\ps{\,   {\rm ps}}
\def\ns{\,   {\rm ns}}
\def\us{\,\mu{\rm  s}}
\def\ms{\,   {\rm ms}}
\def\nA{\,   {\rm nA}}
\def\uA{\,\mu{\rm  A}}
\def\mA{\,   {\rm mA}}
\def\A {\,   {\rm  A}}
\def\mV{\,   {\rm mV}}
\def\V {\,   {\rm  V}}
\def\fF{\,   {\rm fF}}
\def\pF{\,   {\rm pF}}
\def\GeV{\, {\rm GeV}}
\def\MHz{\, {\rm MHz}}
\def\uW{\,\mu{\rm  W}}
\def\e {\,  {\rm e^-}}

\renewcommand{\labelenumi}{\arabic{enumi}}
\renewcommand{\labelitemi}{-}

\begin{frontmatter}

\title{Pixel Detectors for Tracking and their Spin-off in Imaging Applications\thanksref{BMBF} }

\thanks[BMBF]{Work supported by the German Ministerium f{\"u}r Bildung,
              und Forschung (BMBF) under contract
              no.~$05 HA1PD1/5$\ , by the
              Ministerium f{\"u}r Wissenschaft und Forschung (MWF) des Landes
              Nordrhein--Westfalen under contract no.~$IV\,A5-106\,011\,98$, and
              by the DIP Foundation under contract no. E7.1}

% author{Name\corauthref{cor1}\thanksref{label2}}
\author {N.~Wermes\thanksref{NW}}

\thanks[NW]{Physikalisches Institut, Nussallee 12,
               D-53115 Bonn, Germany, Tel.: +49\,228\,73-3533, Fax:
               -3220, email: wermes@physik.uni-bonn.de
           }
\address{Physikalisches Institut der Universit{\"a}t Bonn, Germany}
%\ead{wermes@physik.uni-bonn.de}
% \ead[url]{www.physik.uni-bonn.de/~wermes}

% \hfill\break
% \begin{center}
% Talk given at the 9th European Symposium on Semiconductor
% Detectors, Elmau, Germany, June 2002
% \end{center}

\begin{abstract}
To detect tracks of charged particles close to the interaction
point in high energy physics experiments of the next generation
colliders, hybrid pixel detectors, in which sensor and read-out IC
are separate entities, constitute the present state of the art in
detector technology. Three of the LHC detectors as well as the
BTeV detector at the Tevatron will use vertex detectors based on
this technology. A development period of almost 10 years has
resulted in pixel detector modules which can stand the extreme
rate and timing requirements as well as the very harsh radiation
environment at the LHC for its full life time and without severe
compromises in performance. From these developments a number of
different applications have spun off, most notably for biomedical
imaging. Beyond hybrid pixels, a number of trends and
possibilities with yet improved performance in some aspects have
appeared and presently developed to greater maturity. Among them
are monolithic or semi-monolithic pixel detectors which do not
require complicated hybridization but come as single sensor/IC
entities. The present state in hybrid pixel detector development
for the LHC experiments as well as for some imaging applications
is reviewed and new trends towards monolithic or semi-monolithic
pixel devices are summarized.
\end{abstract}


\end{frontmatter}

\newpage

\section{Radiation hard hybrid pixel detectors for the LHC}
An almost 10 year long development of pixel detectors for tracking
close to the interaction point at the LHC was needed to develop
pixel detector modules which meet the very high demands on spatial
resolution, timing precision, long term operation performance and,
most importantly, radiation tolerance to doses of as much as 500
kGy. All developments have been based on the \emph{hybrid pixel
technique}, in which sensor and FE-chips are separate parts of the
detector module connected by small conducting bumps applied using
the bumping and flip-chip technology. This technique is the only
one which at present is robust and mature enough to cope with the
above demands. All LHC-collider-detectors ALICE
\cite{ALICE_pix,ALICE_Riedler}, ATLAS
\cite{ATLAS_pix,ATLAS_Gemme}, and CMS \cite{CMS_pix,CMS_Erdmann},
LHCb (for the RICH system) \cite{LHCb_pix} at the LHC, BTeV
\cite{BTEV_pix} at the TEVATRON and the CERN fixed target
experiment NA60 \cite{NA60,NA60_Radermacher}, employ the hybrid
pixel technique to build large scale (up to $\sim$2m$^2$) pixel
detectors. Pixel area sizes are typically 50$\mu$m $\times$
400$\mu$m as for ATLAS or 100$\mu$m $\times$ 150 $\mu$m as for
CMS. The detectors are arranged in cylindrical barrels of 2 to 3
layers and disks covering the forward and backward regions.

\begin{figure}[htb]
\begin{center}
%\rotatebox{270}
%{\includegraphics[width=0.5\textwidth]{MatrixANDPhoto.eps}}
\includegraphics[width=1.0\textwidth]{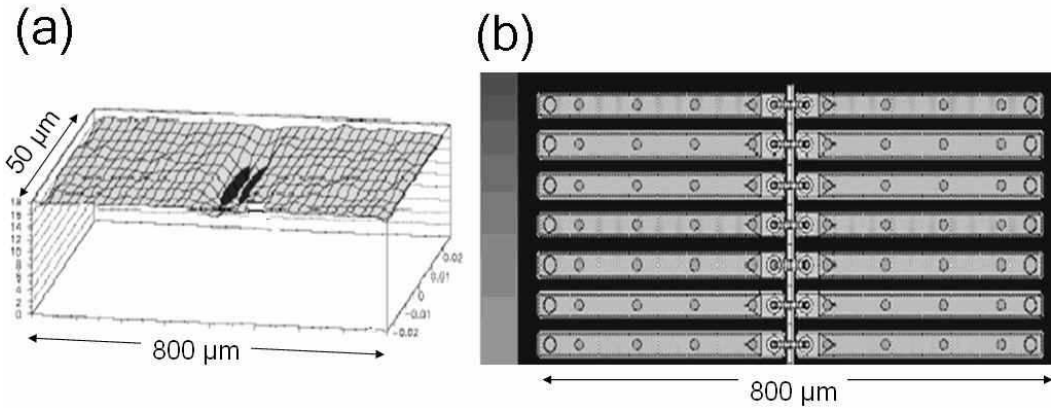}
%\hfill \vspace{0.5cm} \break
%\includegraphics[width=0.20\textwidth]{fig1b.eps}
\end{center}
\caption[]{Charge collection (a) and pixel pattern (b) rows of two
adjacent pixels of the ATLAS pixel detector.} \label{ATLAS-Sensor}
\end{figure}

The discovery that oxygenated silicon is more radiation hard, with
respect to the non-ionizing energy loss of protons and pions,
\cite{oxysilicon} than standard silicon, allows operation of pixel
detectors at the LHC for which the radiation is most severe due to
their proximity to the interaction point. Sensors with $n^+$
electrodes in n-bulk material have been chosen to cope with the
fact that type inversion occurs after about $\Phi_{eq} = 2.5
\times 10^{13}$cm$^{-2}$. After type inversion the $pn$-diode sits
on the electrode side thus allowing the sensor to be operated
partially depleted. Figures \ref{ATLAS-Sensor}(a) and (b) show the
charge collection and the pixel pattern of the ATLAS pixel sensor
\cite{ATLAS_Sensor}. Note the structures of a bias grid in the
center of Fig. \ref{ATLAS-Sensor}(b) which causes a drop of order
10$\%$ in the charge collection efficiency between pixel cells as
shown in Fig. \ref{ATLAS-Sensor}(a). This is considered tolerable
in light of the benefit that the bias grid allows to test the
sensors without a readout chip, an essential necessity for the
building of large area detectors.

\begin{figure}[h]
\begin{center}
\includegraphics[width=1.0\textwidth]{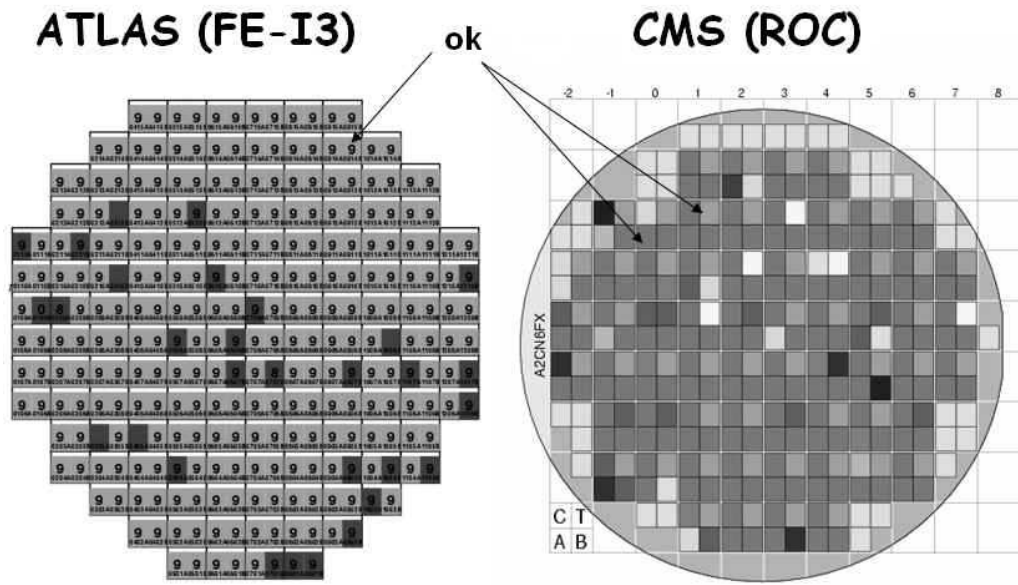}
\end{center}
\caption[]{Wafer maps of the pixel front-end chips of ATLAS and
CMS show high chip yields. For ATLAS the darkly marked chips are
rejects. For CMS the very light and very dark chips are rejected.}
\label{wafermap}
\end{figure}

\begin{figure}[h]
\begin{center}
\includegraphics[width=0.38\textwidth]{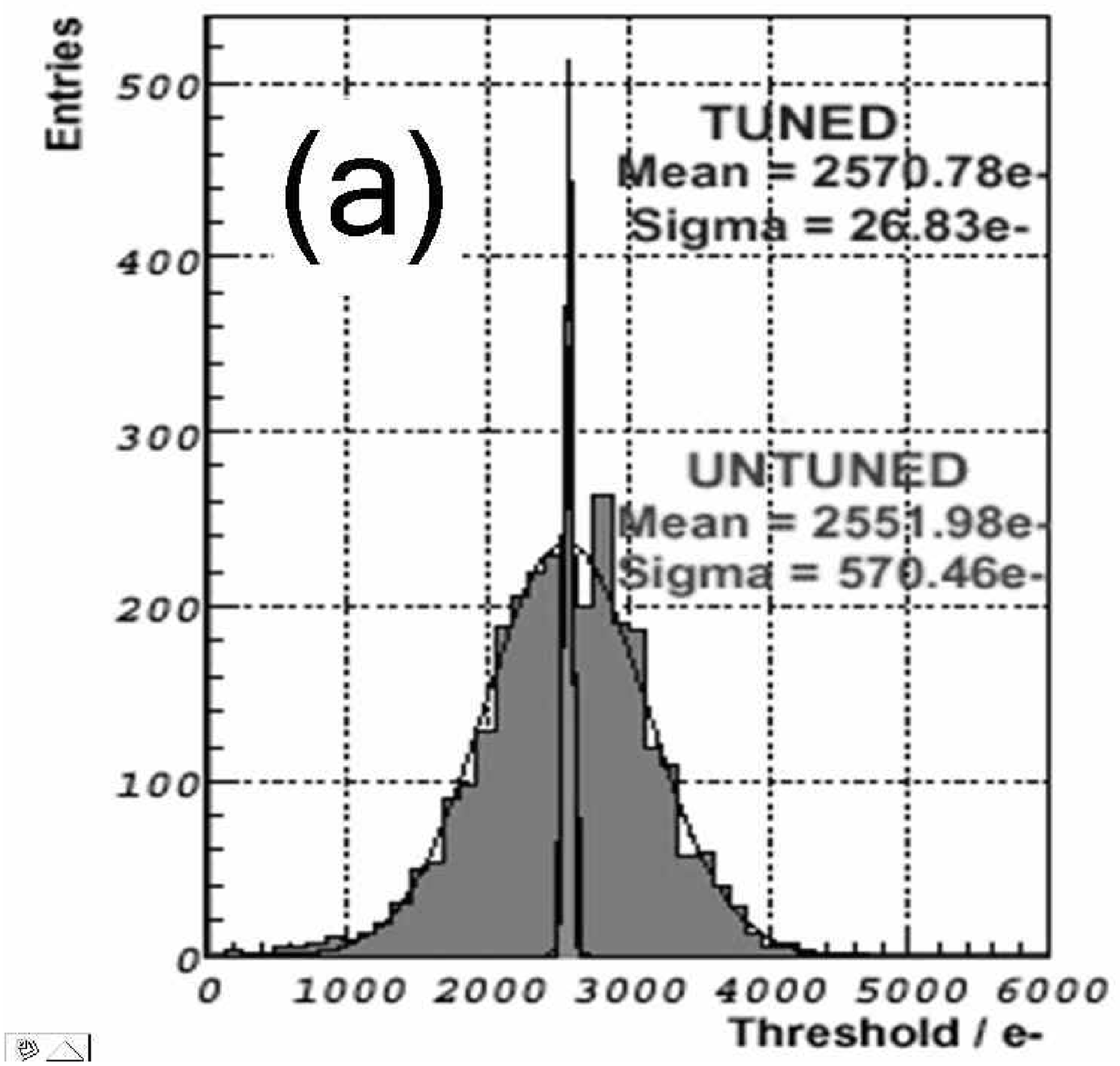}
%\hfill \vspace{0.5cm} \break
\includegraphics[width=0.60\textwidth]{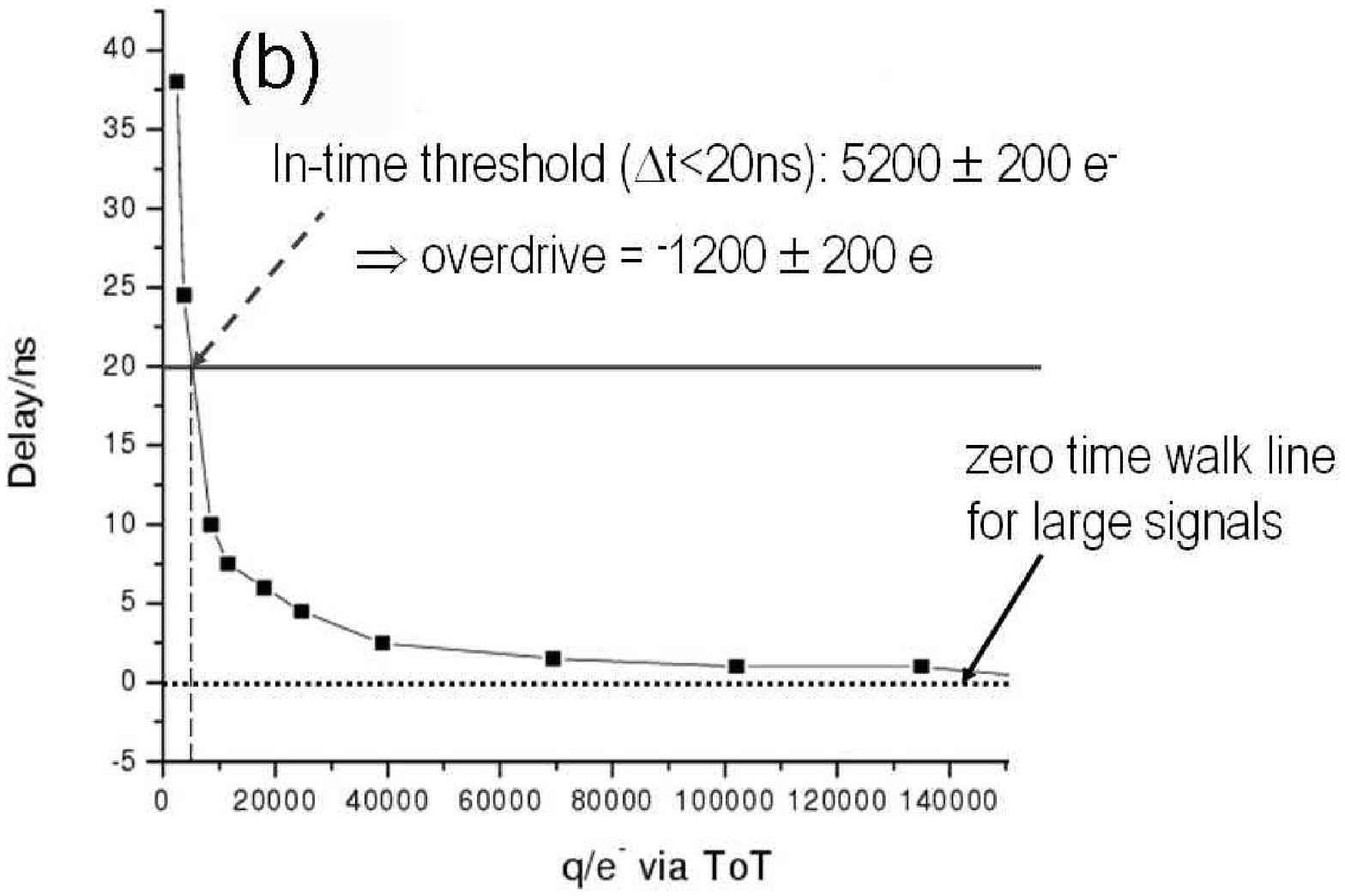}
\end{center}
\caption[]{(a) Dispersion of the pixel thresholds before and after
tuning. (b) In-time threshold and overdrive for a typical
threshold setting of 4000$\pm$200$e^-$.} \label{thresholds}
\end{figure}

The challenge in the design of the front-end pixel electronics
\cite{blanquart03} can be summarized by the following
requirements: low power ($<$ 50$\mu$W per pixel), low noise and
threshold dispersion (together $<$ 200e), zero suppression in
every pixel, on-chip hit buffering, and small time-walk to be able
to assign the hits to their respective LHC bunch crossing. The
pixel groups at the LHC have reached these goals in several design
iterations using first radiation-\emph{soft} prototypes, then
dedicated radhard designs, and finally using deep submicron
technologies. Figure \ref{wafermap} shows two recent wafer maps of
ATLAS and CMS which demonstrate that radiation hard front-end
chips with high yields in excess of $\sim$80$\%$ have been
fabricated for the production of these LHC pixel detectors. While
CMS uses analog readout of hits up to the counting house, ATLAS
obtains pulse height information by means of measuring the
\emph{time over threshold} (ToT) for every hit. Figure
\ref{thresholds}(a) shows the distribution of measured thresholds
of an ATLAS front-end chip. The dispersion of about 600 e$^-$ can
be lowered to below 50 e$^-$ by a 7-bit tuning feature implemented
in the chip. Figure \ref{thresholds}(b) illustrates the effect of
time walk for small signals. For efficient signal detection within
a defined time of 20 ns with respect to the bunch crossing an {\it
overdrive} of about 1200$e^-$ is necessary. The bunch crossing
occurs every 25 ns.

\begin{figure}[htb]
\includegraphics[width=0.36\textwidth]{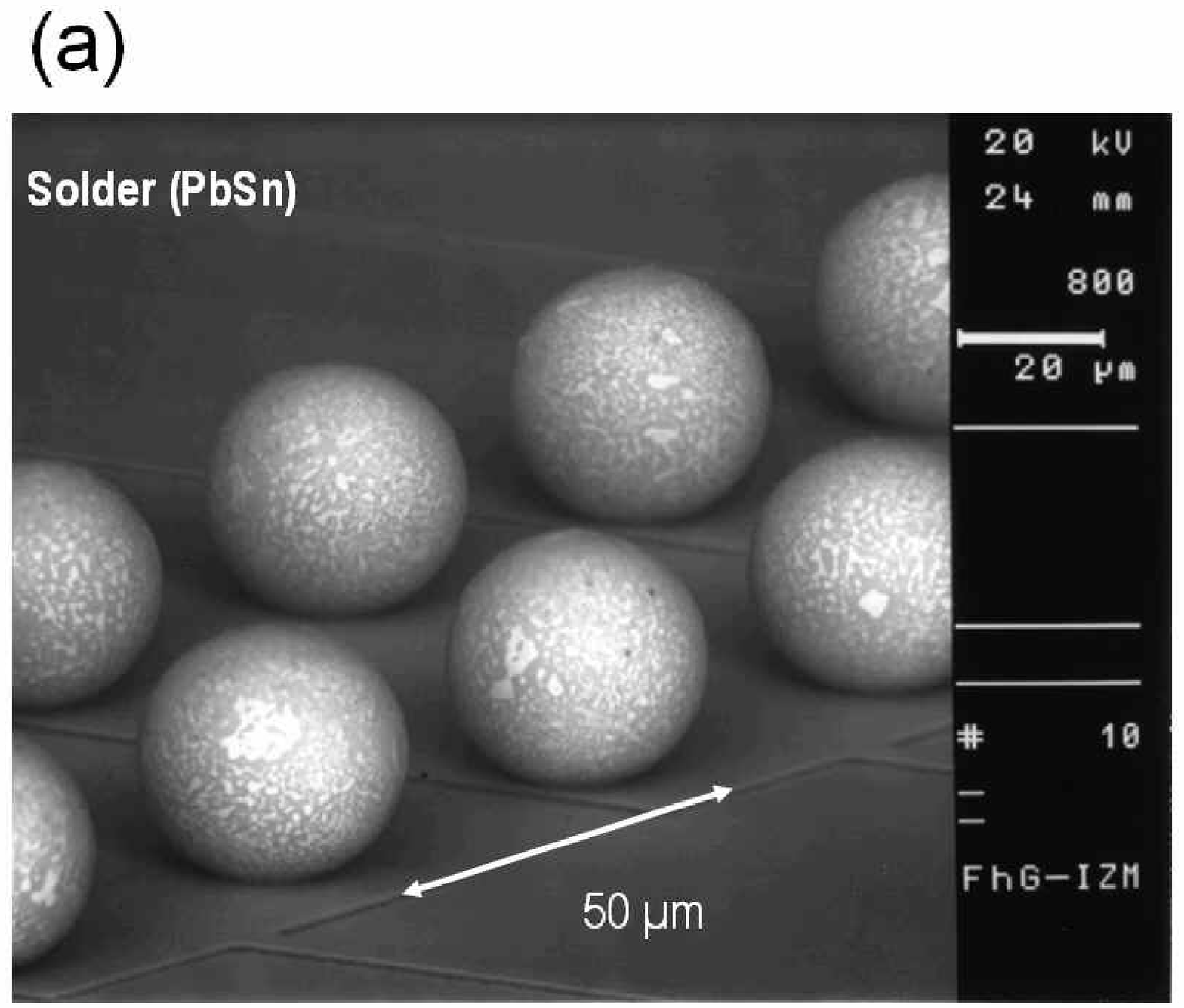}
\includegraphics[width=0.4\textwidth]{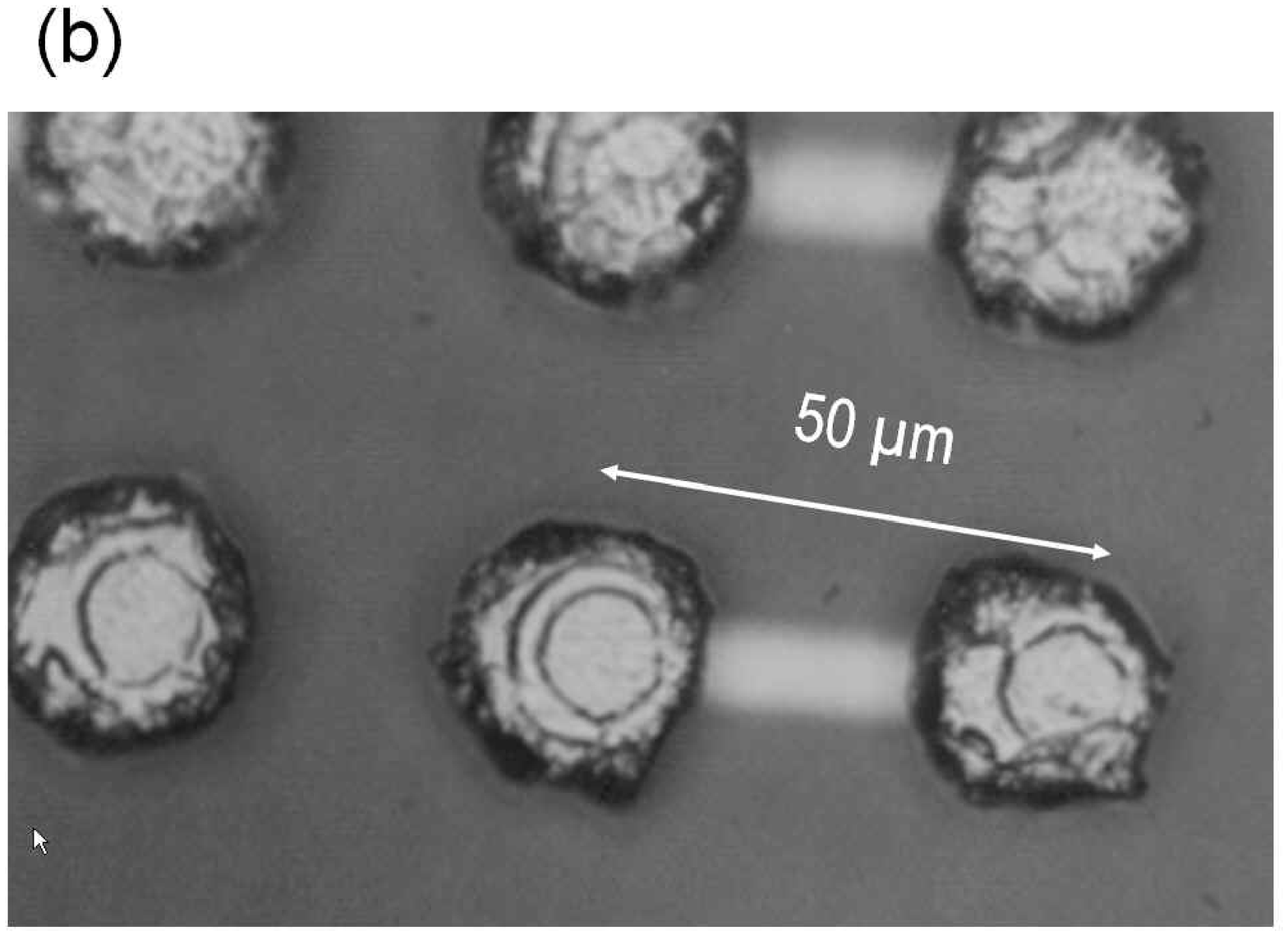}
\includegraphics[width=0.18\textwidth]{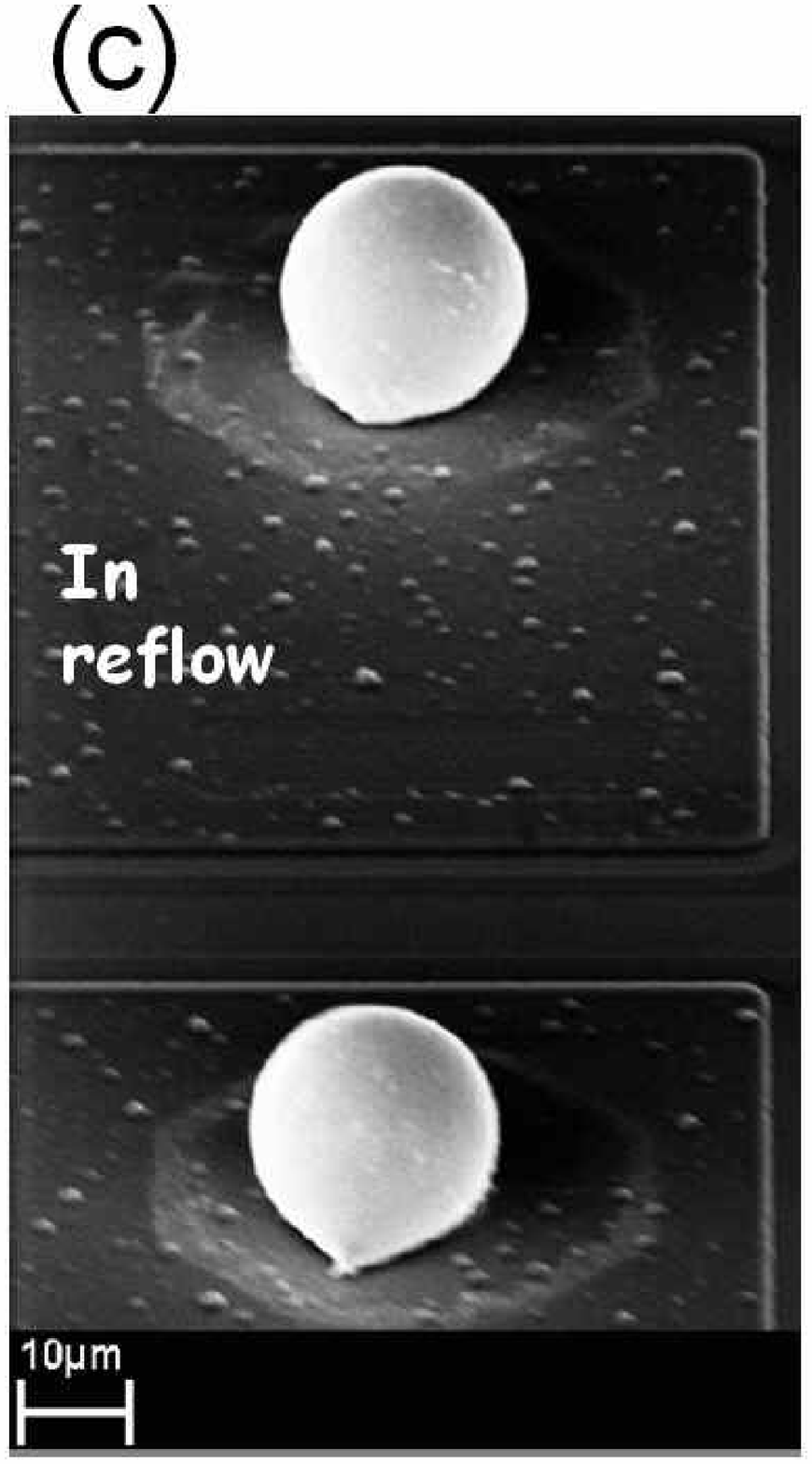}
\caption{(a) solder (PbSn, Photo IZM, Berlin) (b) Indium (Photo
AMS, Rome), and (c) Indium with reflow (Photo PSI, Villigen) bump
rows with $50 \mu$m pitch.} \label{bumps}
\end{figure}

The chip and sensor connection is done by fine pitch bumping and
subsequent flip-chipping which is achieved with either PbSn
(solder) or Indium bumps at a failure rate of $\lesssim 10^{-4}$.
The Indium bumps are applied by a wet lift-off technique and can
be mated by direct thermo-compression
\cite{Indium_bumping,fiorello} or reflowed, as developed by CMS
\cite{In-reflow_bumping}. After bumping the chips are thinned by
backside grinding to a thickness of 150 - 180 $\mu$m. Fig.
\ref{bumps} shows rows of $50 \mu$m pitch bumps obtained by these
techniques. All of these bump bonding technologies have been
successfully used with 8" IC-wafers and 4" sensor wafers.

\begin{figure}[htb]
\includegraphics[width=0.45\textwidth]{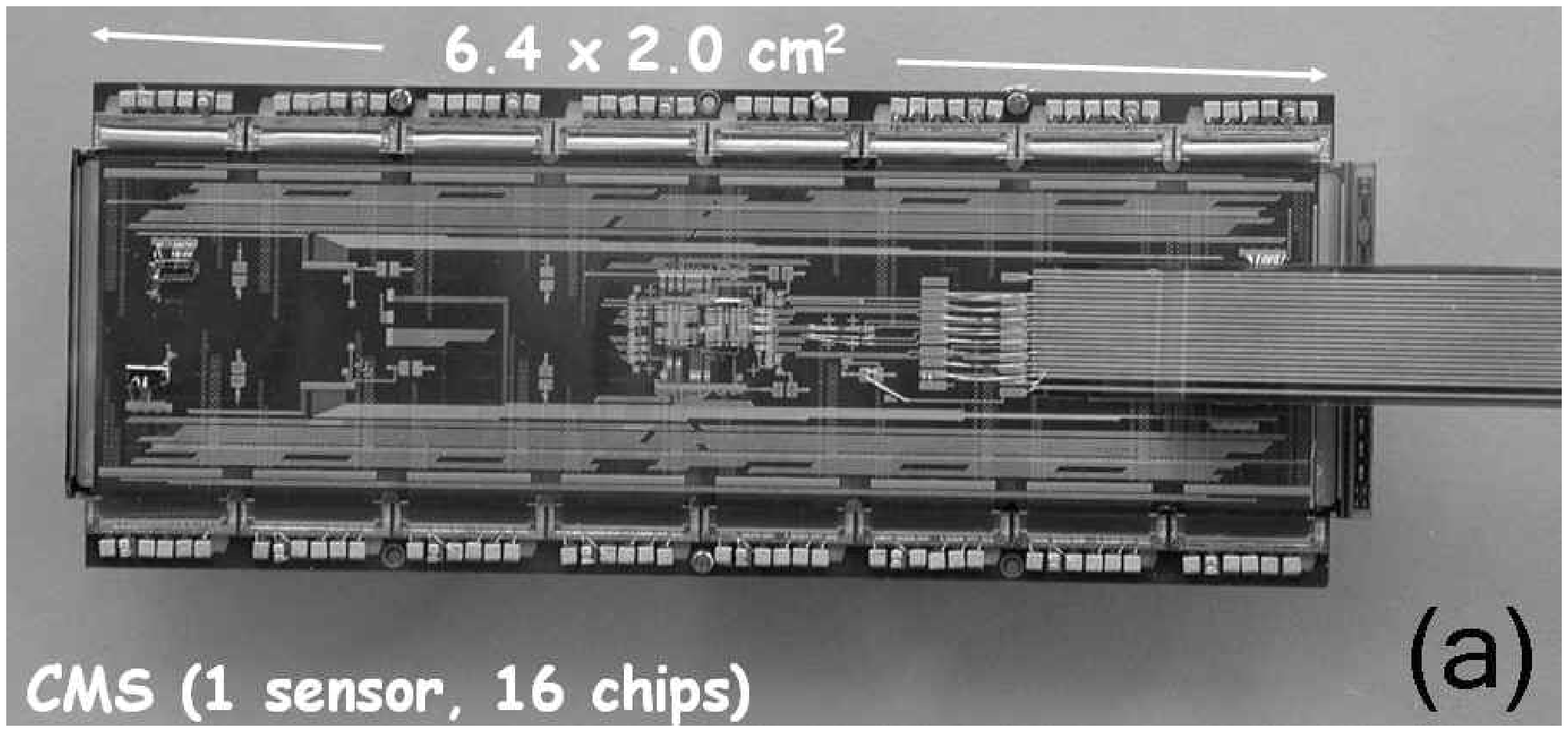}
\includegraphics[width=0.53\textwidth]{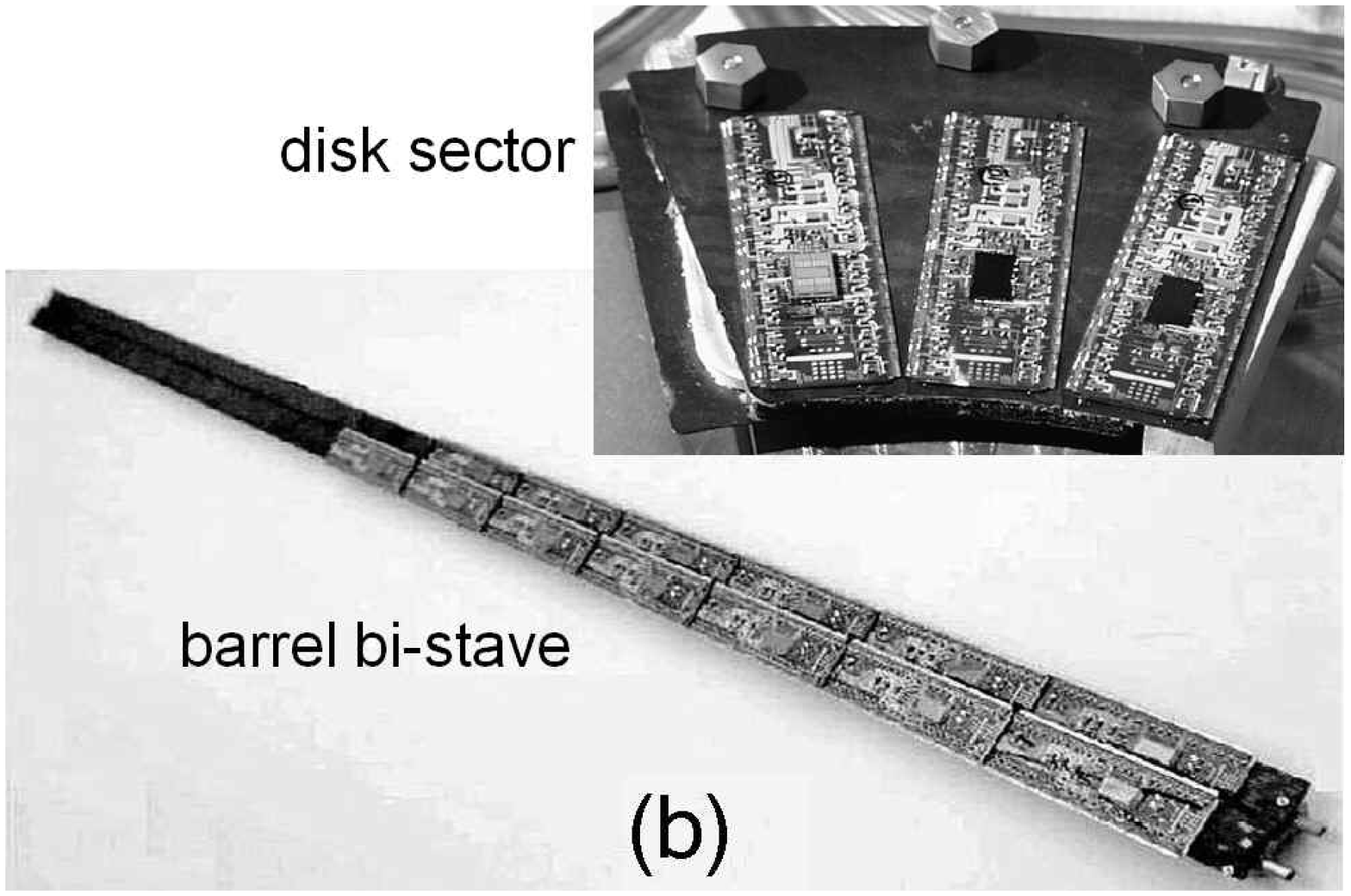}
  \caption{(a) Assembled CMS-pixel module with one sensor and 16 readout chips.
  (b) ATLAS modules mounted to a bi-stave unit and to a disk sector.}
  \label{modules}
\end{figure}

In the case of CMS and ATLAS a \emph{module} (cf Fig.
\ref{modules}(a)) of typically 2 cm $\times$ 6.5 cm area consists
of 16 FE-chips bump-connected to one silicon sensor.The I/O lines
of the chips are connected via wire bonds to a kapton flex circuit
glued atop the sensor. The flex houses a module control chip
responsible for front end time/trigger control and event building.
The total thickness at normal incidence is in excess of $2 \%$
$X_0$. The modules are arranged in barrel-ladders or disk-sectors
as shown in Fig. \ref{modules}(b).

\begin{figure}[htb]
\begin{center}
\includegraphics[width=0.30\textwidth]{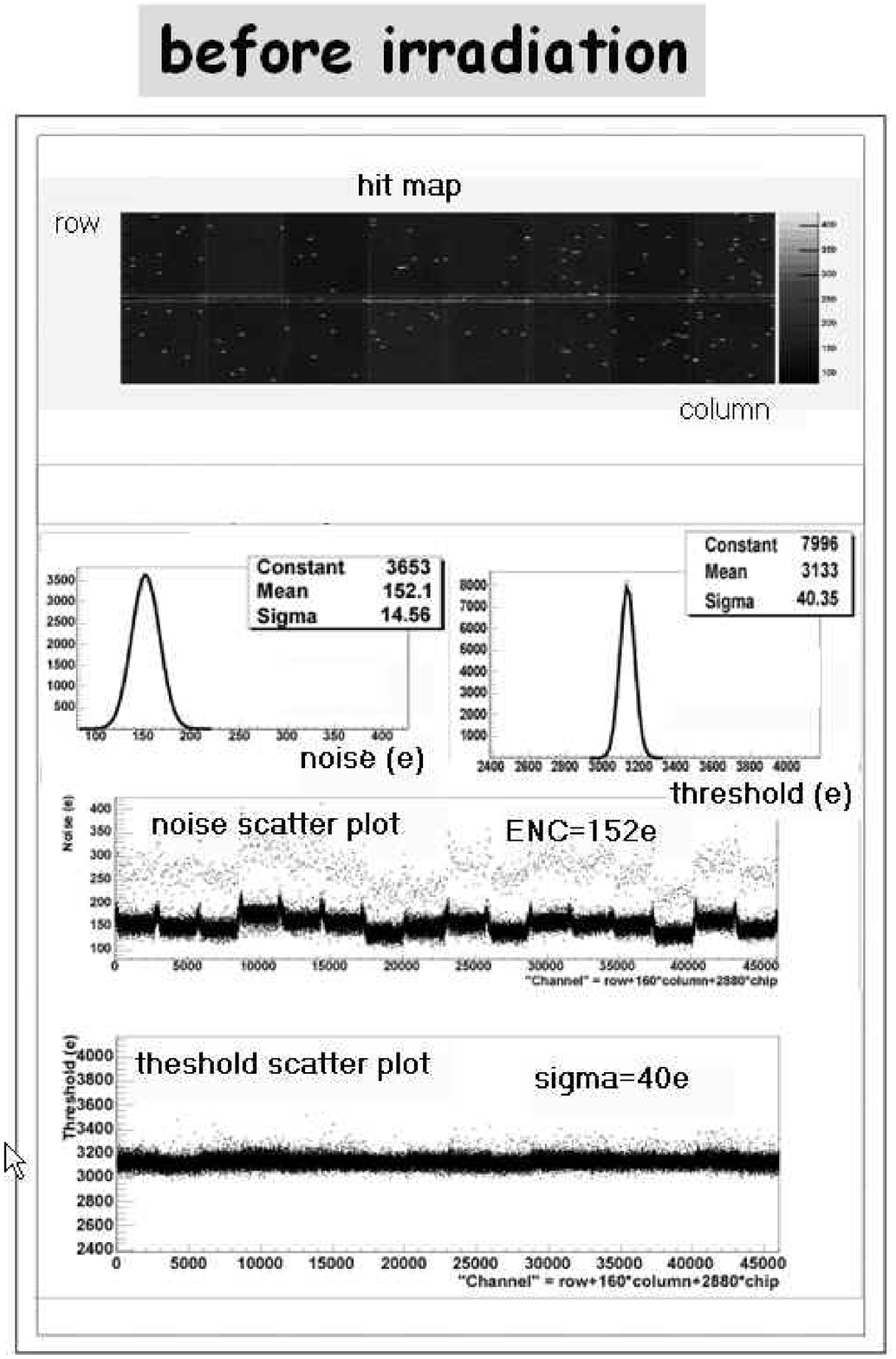}
\includegraphics[width=0.31\textwidth]{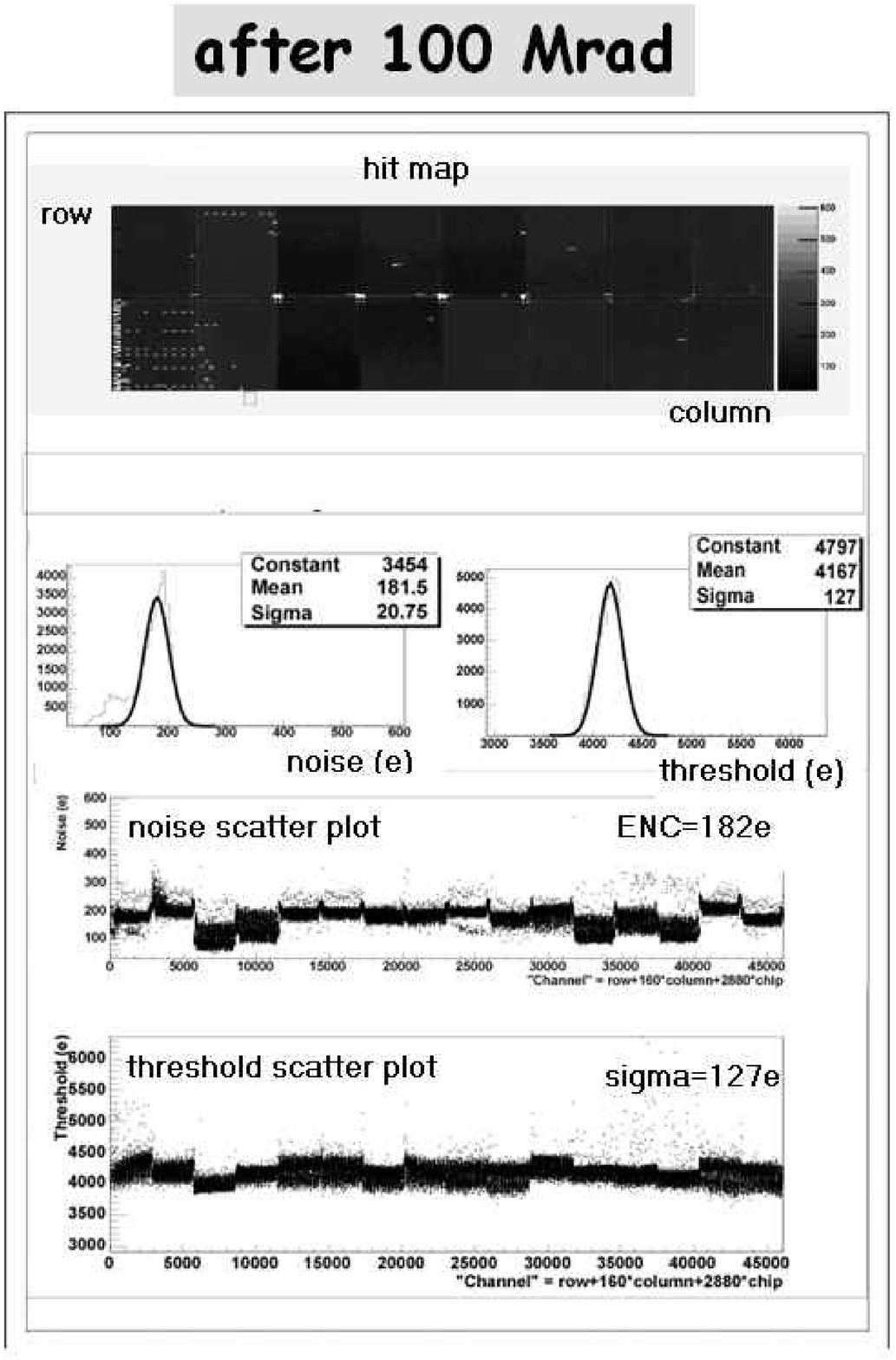}
\vskip 0.5cm
\includegraphics[width=0.65\textwidth]{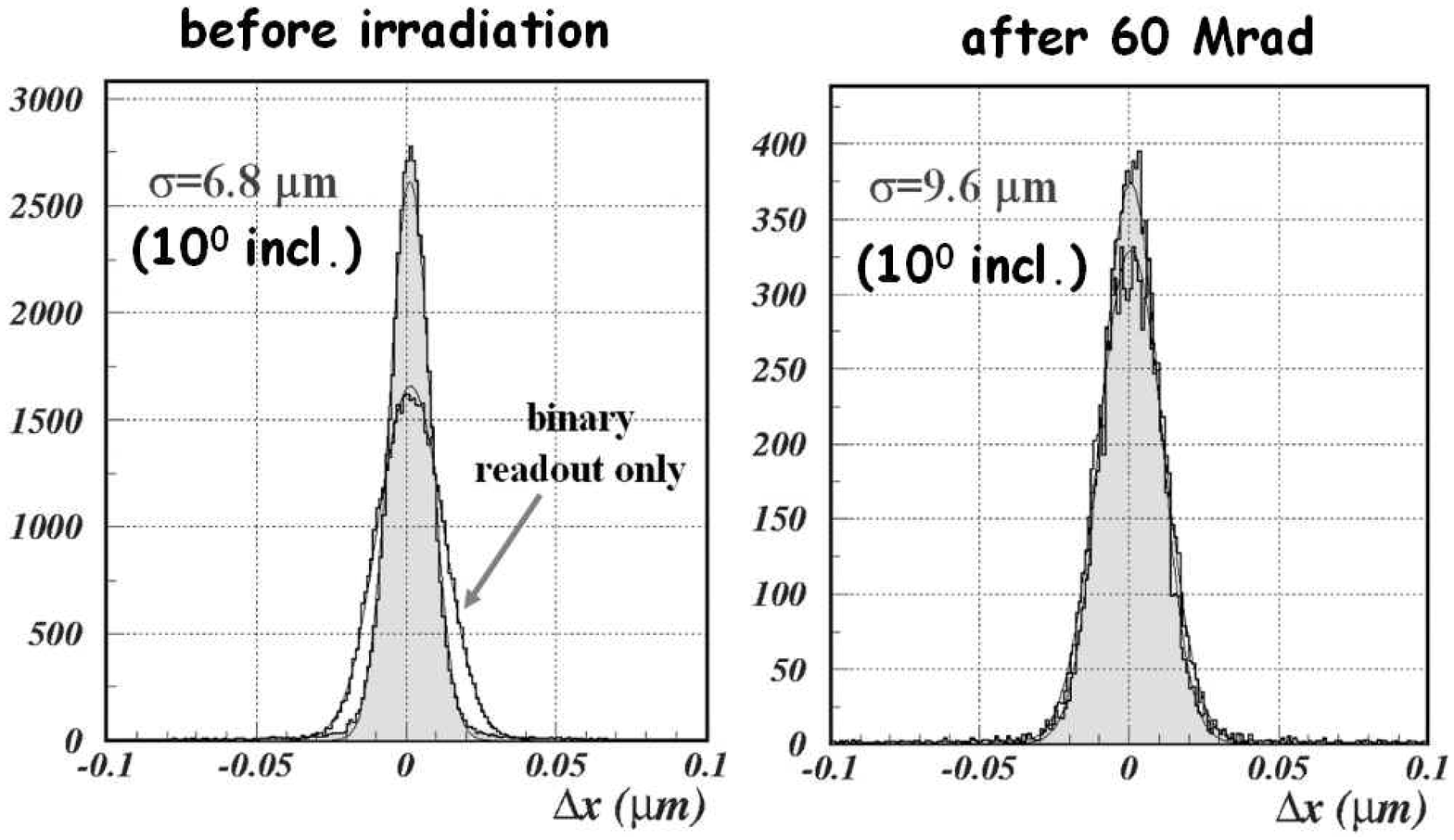}
\vskip 0.3cm
\includegraphics[width=0.65\textwidth]{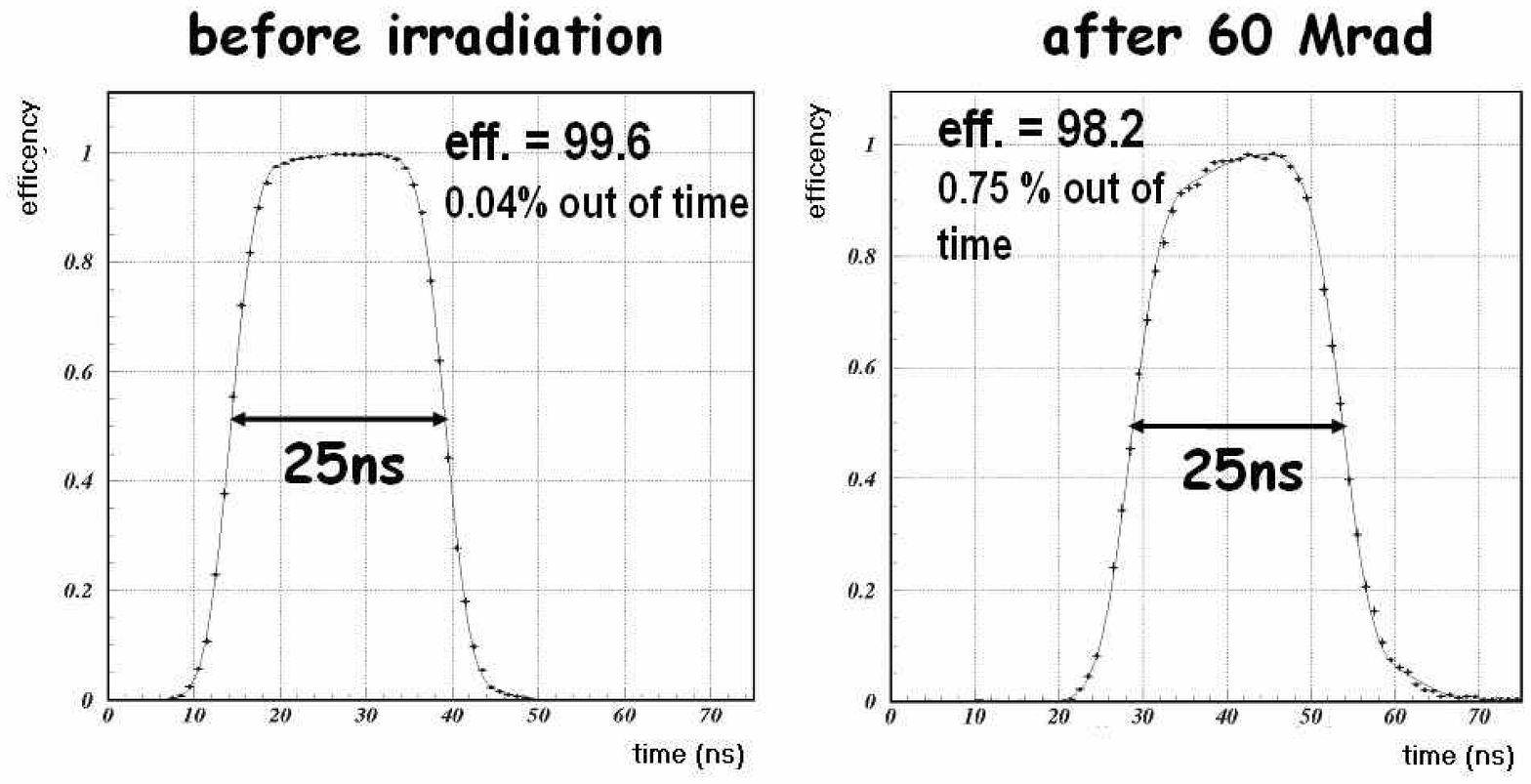}
\end{center}
  \caption{Comparisons of ATLAS pixel modules before and after
  irradiation to doses up to 100 Mrad.
  Hit map, noise and threshold dispersions (top), spatial
  resolution in the 50 $\mu$m direction of the pixels at 10$^o$
  incidence angle (center), and
  the hit efficiency (bottom). The in-time efficiency of a hit to be earlier than 25 ns
  is determined
  in test beams relative to a fixed delay of the trigger counters.
  The highest point of the plateau shows the in-time efficiency. The width
  of the plateau characterizes the available margin during operation.}
  \label{irrad}
\end{figure}

The most challenging task in the pixel development for LHC was to
meet the very high radiation dose that a pixel detector is exposed
to during 10 years of operation (500 kGy). The advancement of
oxygenated silicon and deep submicron chip technology made a long
life time in such an environment possible. Figures
\ref{irrad}(a)-(c) show the comparison of critical performance
figures before and after irradiation of ATLAS pixel modules. In
parts the received dose of the modules was well in excess of that
expected for 10 years operation at the LHC. After irradiation to
500 kGy the mean collected charge fraction has been measured to be
87$\pm$14$\%$ and the in-time efficiency, i.e. the efficiency for
hit detection within 20ns after the bunch crossing, is 98.2$\%$.

As the physics to be addressed with ALICE pixels is different from
those addressed by ATLAS and CMS, a pixel module for ALICE has
different key features. Because in ALICE the expected radiation is
reduced ($< 5$kGy, $6\times 10^{12}$n$_{eq}$) cooling to
temperatures below 0$^\circ$C as for CMS and ATLAS is not
mandatory. Very low total radiation length values are hence aimed
for and achieved by the reduced cooling requirements (to
24$^\circ$ C), thus contributing only with 0.1$\%$X$_0$, by a very
light weight carbon fibre support structure (0.1$\%$X$_0$), and by
using thin sensors (200 $\mu$m) and chips (150 $\mu$m). An ALICE
module (1.28 cm $\times$ 7.0 cm) contains 5 readout chips bonded
to one sensor. The total radiation length per layer is only
0.9$\%$.

%The modules are positioned by dedicated robots on carbon-carbon
%ladders (staves) and cooled by evaporation of a fluorinert liquid
%($C_3F_8$) at an input temperature below $-20 ^{o}C$ in order to
%maintain the entire detector below $-6 ^{o}C$ to minimize the
%damage induced by radiation. This operation requires pumping and
%the cooling tubes must stand $16$ bar pressure if pipe blocking
%occurs. All detector components must survive temperature cycles
%between $-25 ^{o}C$ and room temperature.

\begin{figure}[htb]
\begin{center}
\includegraphics[width=0.9\textwidth]{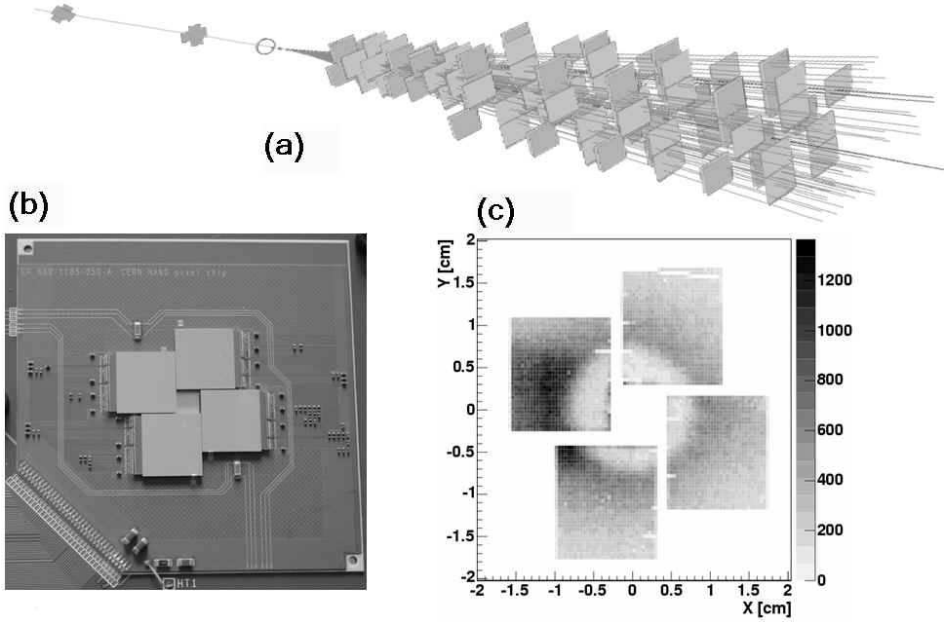}
\end{center}
  \caption{Pixel detector tracker in the NA60 experiment (a) consisting of
    eight 8-chip planes and eight 4-chip planes (b).  The pixel planes
    were operated through partial type-inversion, with the results
    demonstrated in the hit-multiplicity plot (c).
  }
  \label{NA60}
\end{figure}

The CERN heavy ion experiment NA60 \cite{NA60_Radermacher} has
used LHC-type pixel detectors for the first time in a running
experiment. The setup of the NA60 pixel tracker is shown in Fig.
\ref{NA60}(a). For the initial running the ALICE-LHCb chip was
used. Eight 4-chip (Fig. \ref{NA60}(b)) and eight 8-chip planes
provide track reconstruction with 11 pixel hits on a track. In a
recent run the sensors have been exposed to a radiation dose of
1.2 kGy and were operated through type inversion. Due to the
inhomogeneous irradiation the inner part of the planes has
received a larger dose than the outer, which is demonstrated by
the hit multiplicity pattern taken with a lowered bias voltage in
Fig. \ref{NA60}(c). The NA60 experiment has now upgraded the pixel
detector by using original ATLAS pixel production modules.

\section{Imaging with Hybrid Pixel Detectors}
\begin{figure}[htb]
\begin{center}
\includegraphics[width=0.45\textwidth]{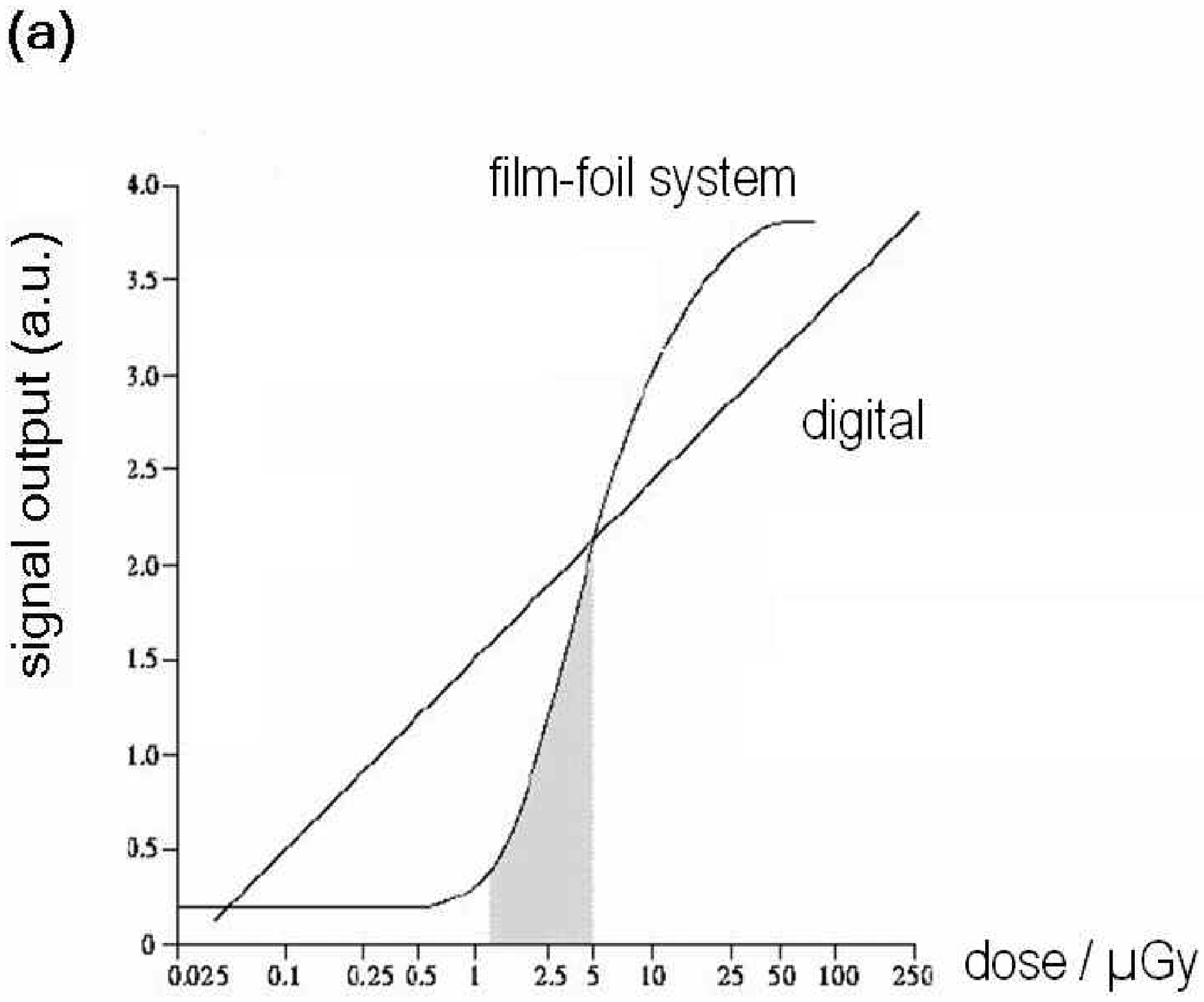}
\includegraphics[width=0.45\textwidth]{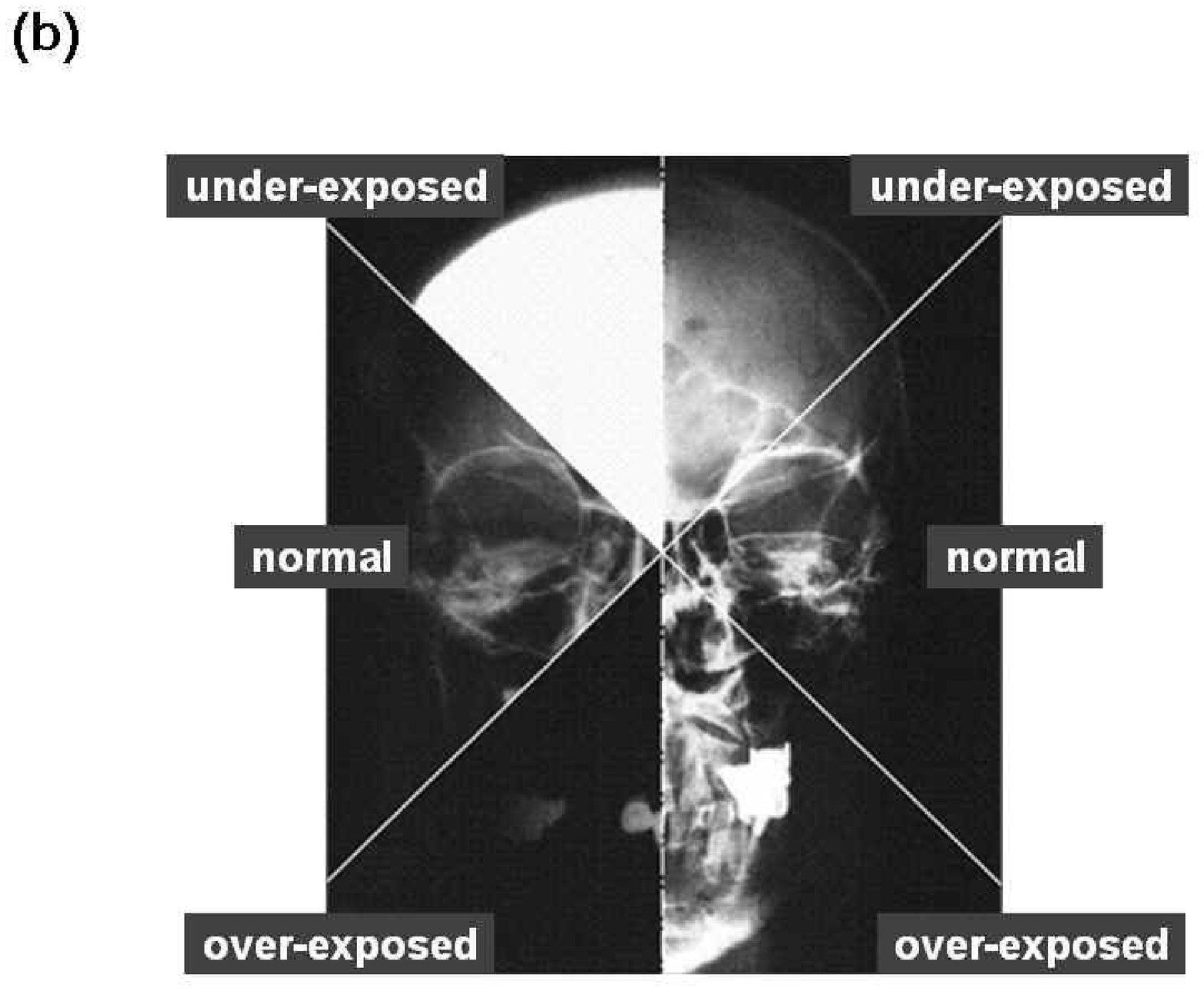}
\end{center}
  \caption{In comparison to X-ray imaging with film-foil systems imaging by counting can offer a linear response (a) and
  a very good image contrast without over and under exposure (b).
  }
  \label{Xray}
\end{figure}

Hybrid pixel detectors have had an impact on imaging applications
as detectors that accumulate the incident radiation through the
counting of individual radiation quanta in every pixel cell. This
technique offers many features which are very attractive for X-ray
imaging: full linearity in the response function, in principle an
infinite dynamic range, optimal exposure times and a good image
contrast compared to conventional film-foil based radiography (cf.
Fig. \ref{Xray}). It thus avoids over- and under-exposed images.
Counting pixel detectors must be considered as a very direct
spin-off of the detector development for particle physics into
biomedical applications. The analog part of the pixel electronics
is in parts close to identical to the one for LHC pixel detectors
while the periphery has been replaced by counting circuitry
\cite{PeFicounter}. The same principle is also used for
protein-crystallography with synchrotron radiation
\cite{Graafsma_Portland,3Dwestbrook}.

The challenges which are to be addressed in order to be
competitive with integrating systems are: high speed ($>$ 1 MHz)
counting with a range of at least 15 bits, operation with very
little dead time, low noise and particularly low threshold
operation with small threshold dispersion values. In particular
the last item is important in order to allow homogeneous imaging
of soft X-rays of energies in the energy range below 10 keV. It is
also mandatory for a differential energy measurement, realized so
far as a double threshold with energy windowing logic
\cite{MPEC-windowing1,MPEC-windowing2,MEDIPIX2}, which can enhance
the contrast of an image as the shape of the X-ray energy spectrum
is different behind different absorbers (e.g. bone or soft
tissue). Finally, for radiography, high photon absorption
efficiency is mandatory, requiring the use and development of high
Z sensors and their hybridization.

\begin{figure}
\begin{center}
\includegraphics[width=0.9\textwidth]{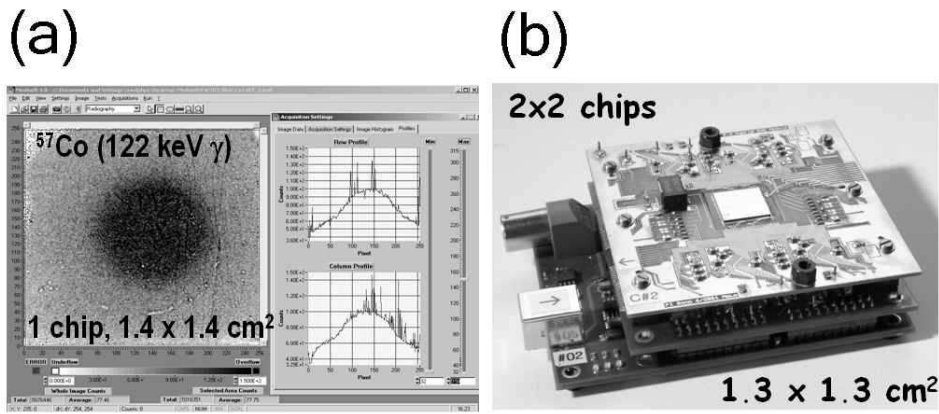}
\end{center}
\caption[]{(a) Image of a $^{57}$Co (122 keV $\gamma$) point
source taken with a MEDIPIX2 counting single chip module (14x14
mm$^2$)\cite{MEDIPIX_Portland}, and (b) MPEC 2x2 multi chip module
with a CdTe sensor \cite{MPEC_Portland}.} \label{counting}
\end{figure}

Several counting pixel system development efforts are being
carried out: the MEDIPIX collaboration \cite{MEDIPIX,MEDIPIX2}
uses the MEDIPIX chip with $256 \times 256$, 55 $\times $55$
\mu$m$^2$ pixels fabricated in a $0.25 \mu$m technology, energy
windowing via two tunable discriminator thresholds, and a 13 bit
counter. The maximum count rate per pixel is about 1 MHz. Fig.
\ref{counting}(a) shows an image of a $^{57}$Co (122 keV $\gamma$)
1 mm diameter point source obtained with the Medipix2 SINGLE chip
bonded to a 14x14 mm$^2$ CdTe sensor \cite{MEDIPIX_Portland}. A
Multi-Chip module with 2x2 chips using high-Z CdTe sensors with
the MPEC chip \cite{MPEC_Portland} is shown in Fig.
\ref{counting}(b). The MPEC chip features $32 \times 32$ pixels
(200$\times$200$\mu$m$^2$), double threshold operation, 18-bit
counting at $\sim$1 MHz per pixel as well as low noise values
($\sim$120e with CdTe sensor) and threshold dispersion ($21$e
after tuning) \cite{MPEC_ref,MPEC_Portland}. A technical issue
here is the bumping of individual die CdTe sensors which has been
solved using Au-stud bumping with In-filling \cite{MPEC_CdTe}. A
challenge for counting pixel detectors in radiology is to build
large area detectors. Commercially available integrating
pixellated systems such as flat panel imagers
\cite{flatpanel1,flatpanel2,flatpanel3,flatpanel4,a-Se} set the
competition level.

\begin{figure}
\begin{center}
\includegraphics[width=0.65\textwidth]{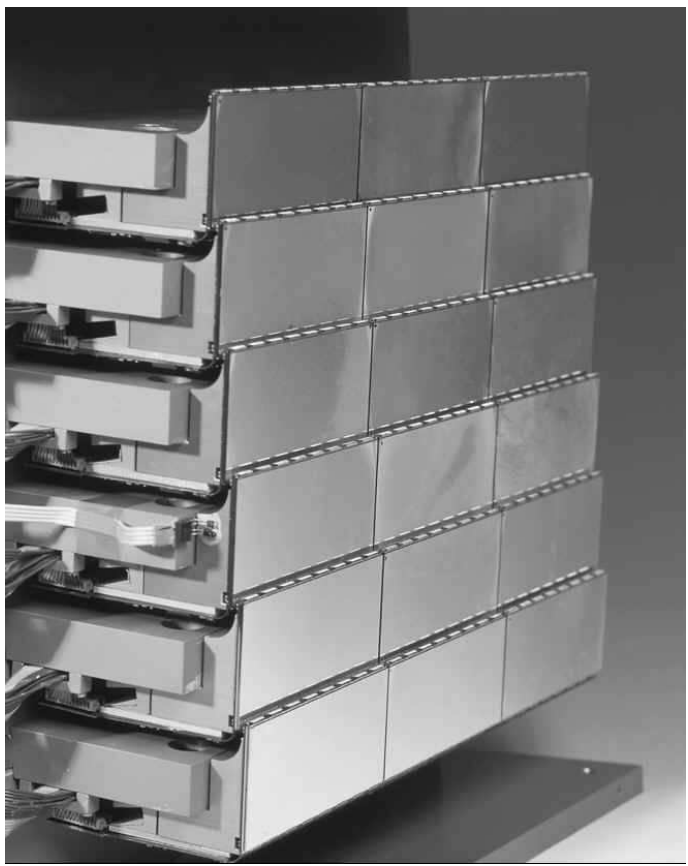}
\vskip 0.5cm
\includegraphics[width=0.50\textwidth]{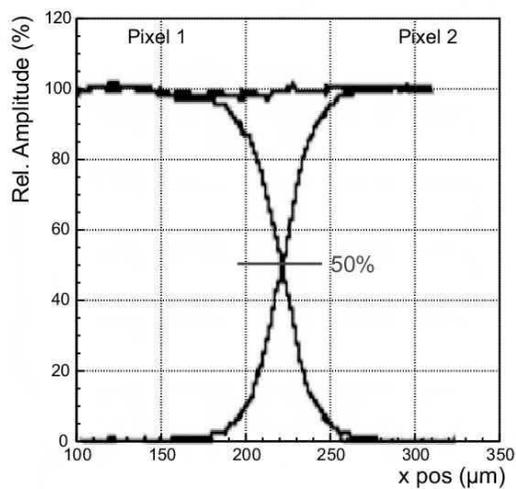}
\hskip 0.5cm
\includegraphics[width=0.45\textwidth]{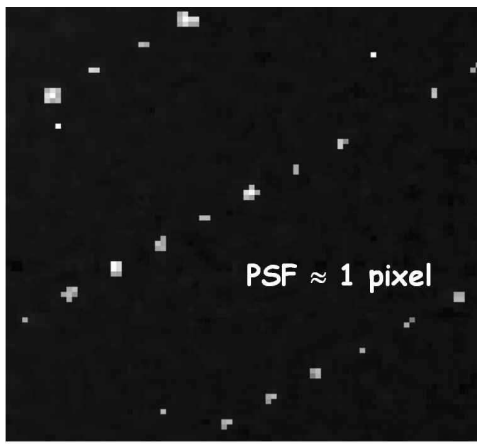}
\end{center}
\caption[]{(top) Photograph of the 20x24 cm$^2$ large PILATUS 1M
detector for protein crystallography using counting hybrid pixel
detector modules, (bottom left) delicate threshold tuning with
counting pixel detectors at the borders in between pixels, (bottom
right) Bragg spots of an image of Lysozyme taken with PILATUS 1M
\cite{PILATUS_Portland} are often contained in one pixels.}
\label{crystallography}
\end{figure}

\hfill\break In {\it protein crystallography} with synchrotron
radiation \cite{Graafsma_Portland} the challenge is to image many
thousands of Bragg spots from X-ray photons with energies of
$\sim$12 keV (corresponding to resolutions at the 1$\AA$ range) or
higher, scattering off protein crystals. This must be accomplished
at a high rate ($\sim$1-1.5 MHz/pixel) and by systems with a high
dynamic range. The typical spot size of a diffraction maximum is
$100-200 \mu$m, calling for pixel sizes in the order of $100-300
\mu$m. The high linearity of the hit counting method and the
absence of so-called "blooming effects", i.e. the response of
non-hit pixels in the close neighborhood of a Bragg spot, makes
counting pixel detectors very appealing for protein
crystallography experiments. A systematic limitation and
difficulty is the problem that homogeneous hit/count responses in
all pixels, also for hits at the pixel boundaries or between
pixels where charge sharing plays a role must be maintained by
delicate threshold tuning (cf Fig. \ref{crystallography}(b)).
Counting pixel developments are made for the ESRF (Grenoble,
France) \cite{XPAD1,XPAD2} and the SLS (Swiss Light Source at the
Paul-Scherrer Institute, Switzerland) beam lines. A photograph of
the PILATUS 1M detector \cite{PILATUS} at the SLS ($\sim$ 10$^6$
$217 \mu$m $\times 217 \mu$m pixels, 18 modules, 20$\times$24
cm$^2$ area) is shown in Fig. \ref{crystallography} (top
photograph). It is the first large scale hybrid pixel detector in
operation. Fig. \ref{crystallography}(c) shows some Bragg spots
obtained from a Lysozyme crystal with a 10s exposure to 12 keV
sychrotron X-rays \cite{PILATUS_Portland}. Some spots are
contained in only one pixel, others spread over a few pixels due
to charge sharing. This demonstrates the intrinsically good point
resolution of the system. Alternative developments which aim to
improve the active/inactive area ratio for protein-crystallography
X-ray detection are so-called 3-D silicon sensors (strip or
pixels) \cite{3D-parker}. A detailed account can be found in
\cite{3D_hiroshima}.

%In summary hybrid pixel detectors are {\underline {the}} present
%state of the art in pixel detector technology. For tracking as
%well as for imaging applications large ($\sim$m$^2$) detectors are
%being built or already in operation (PILATUS). On the negative
%side are the comparatively large material budget, the complicated
%assembling (hybridization) with potential yield pitfalls, and the
%difficulty to obtain high assembly yields with non-wafer scale
%high-Z sensor materials. As trends within this technology,
%interleaved pixels and MCM-D promise better resolution and more
%homogeneous modules, while 3D-silicon detectors can address
%applications in high radiation environments or those where a large
%active/inactive area ratio is mandatory.

\section{New directions with Hybrid Pixel Detectors}
To further develop Hybrid Pixel detectors for use in specialized
applications and to improve in certain areas, several ideas and
developments are being pursued.

\subsection*{Diamond Pixels}
\begin{figure}
\begin{center}
\includegraphics[width=1.0\textwidth]{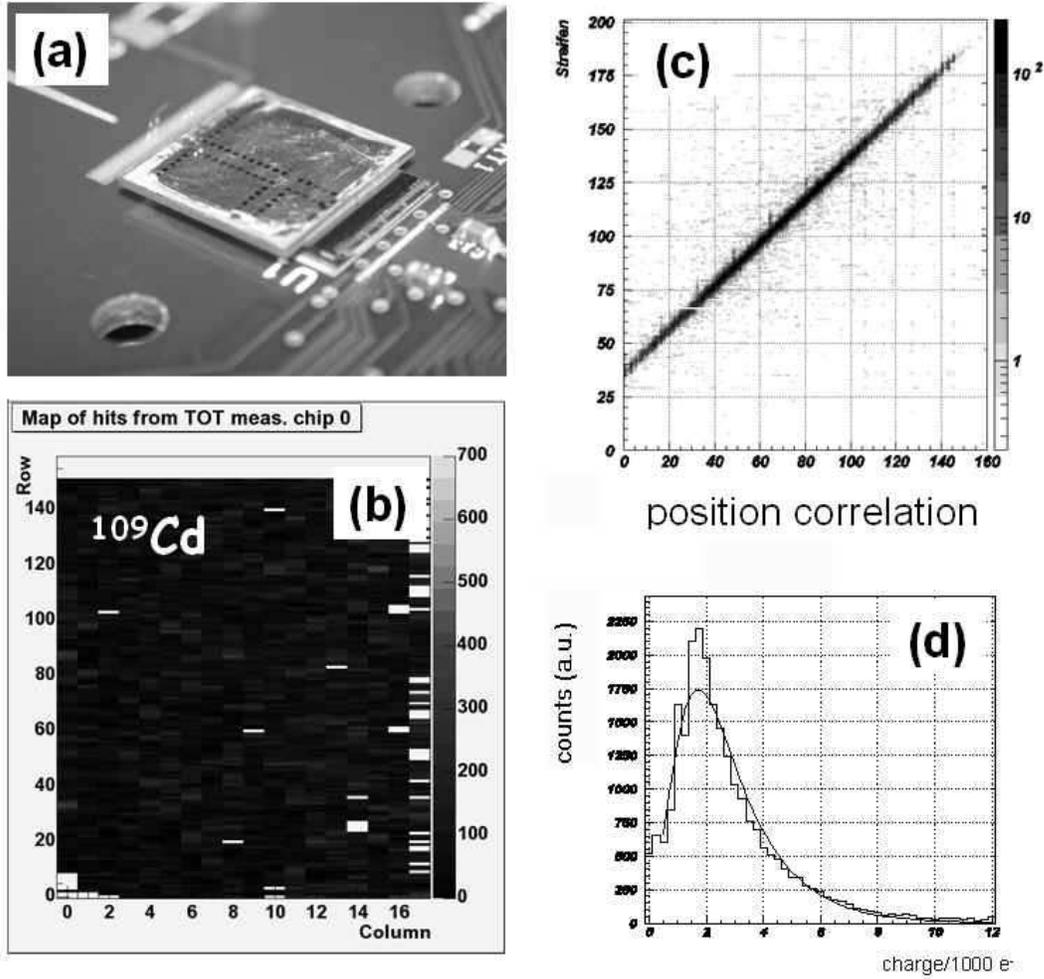}
\end{center}
\caption[]{(a) Single chip diamond pixel module using ATLAS
front-end electronics, (b) hit map obtained by exposure to a
$^{109}$Cd radioactive source (22 keV $\gamma$), (c) scatter plot
of position correlation between the diamond pixel detector and a
reference beam telescope, and (d) measured Landau distribution in
a CVD-diamond pixel detector.} \label{diamond}
\end{figure}

The successful development of radiation hard CVD-diamond sensors
\cite{RD42} with charge collection distances approaching 300
$\mu$m has triggered the developments of a hybrid pixel detector
using diamond as sensors \cite{diamond-pixels}. The non-uniform
field distribution inside CVD-diamond, which originates from the
grain structure in the charge collection bulk, leads to charge
trapping at the grain boundaries \cite{lari04} and -- as a result
-- to shift in the position of reconstructed hits of the order of
100$\mu$m - 150$\mu$m. Diamond pixel detectors are well suited to
study this effect \cite{lari04}. Single chip pixel modules using
ATLAS front-end electronics have been built and tested in a high
energy (180 GeV) pion beam. Figure \ref{diamond}(a) shows the
diamond pixel detector, Fig. \ref{diamond}(b) is a hit response
pattern obtained by exposing the detector to a $^{109}$Cd source
of 22 keV $\gamma$ rays, which deposits approximately 1/4 of the
charge of a minimum ionizing particle. This demonstrates the good
charge collection efficiency obtained. Diamond sensors with charge
collection distances in in excess of 300 $\mu$m have been
fabricated and tested \cite{Kagan_Hiroshima}. Figure
\ref{diamond}(c) and (d) show position correlation and the charge
distribution of the diamond pixel detector in a high energy beam,
respectively. A spatial resolution of $\sigma$ = 22$\mu$m has been
measured with 50$\mu$m pixel pitch, compared to about 12$\mu$m
with Si pixel detectors of the same geometry using the same
electronics, revealing the influence of the mentioned position
shifts.

\subsection*{HAPS}
In order to achieve smaller pixel implant pitches with relaxed
readout cell pitches capacitive coupling between pixels can be
exploited as is done in the HAPS (Hybrid (Active) Pixel Sensors)
concept \cite{HAPS}. This technique is adopted from capacitively
coupled silicon micro strip detectors. The ratio of the number of
interleaved to read-out pixels can be as large as 22/3
\cite{HAPS}. The charge accumulated on interleaved pixels couples
via inter-pixel capacitances to the pixel cells which are read
out. The implant pitch is designed for best spatial resolution
using charge sharing between neighbors, while the readout pitch is
tailored to the size needs of the front-end electronics cell. In
this way resolutions between 3$\mu$m and 10$\mu$m can be obtained
with pixel (readout) pitches of 100$\mu$m (200$\mu$m) \cite{HAPS}.

\subsection*{MCM-D}
\begin{figure}[htb]
\includegraphics[width=0.57\textwidth]{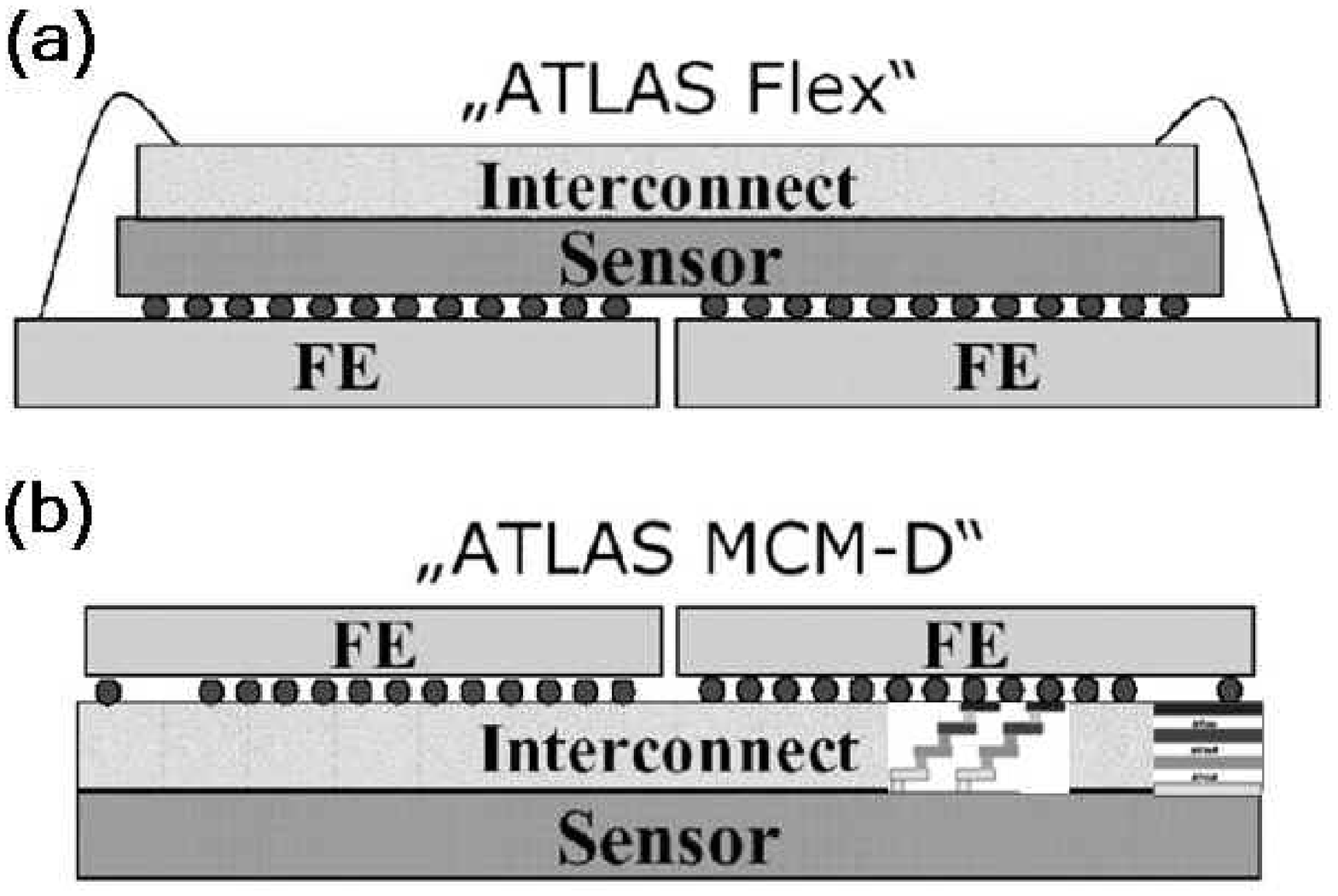}
\includegraphics[width=0.38\textwidth]{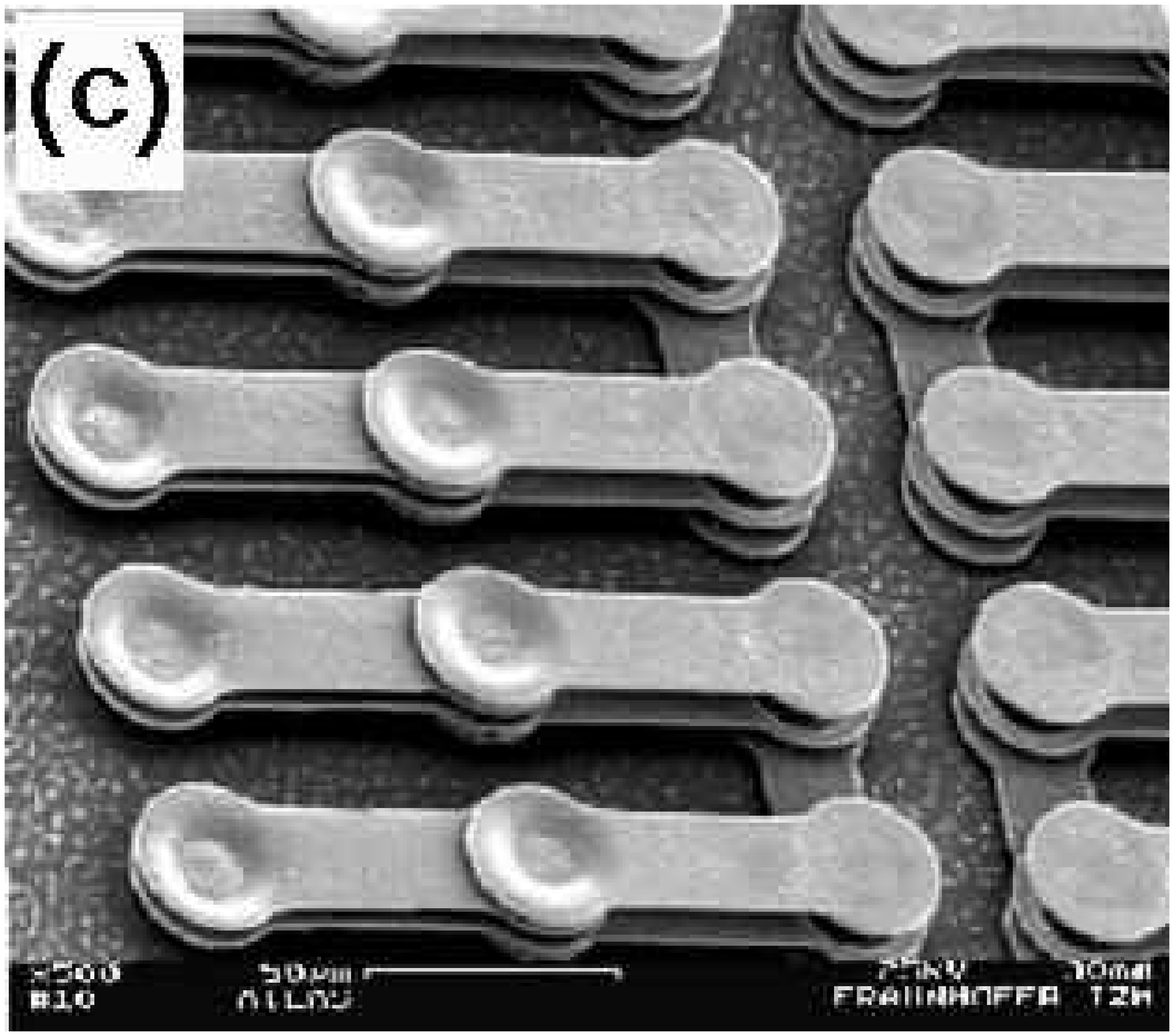}
\vskip 0.5cm
\includegraphics[width=0.95\textwidth]{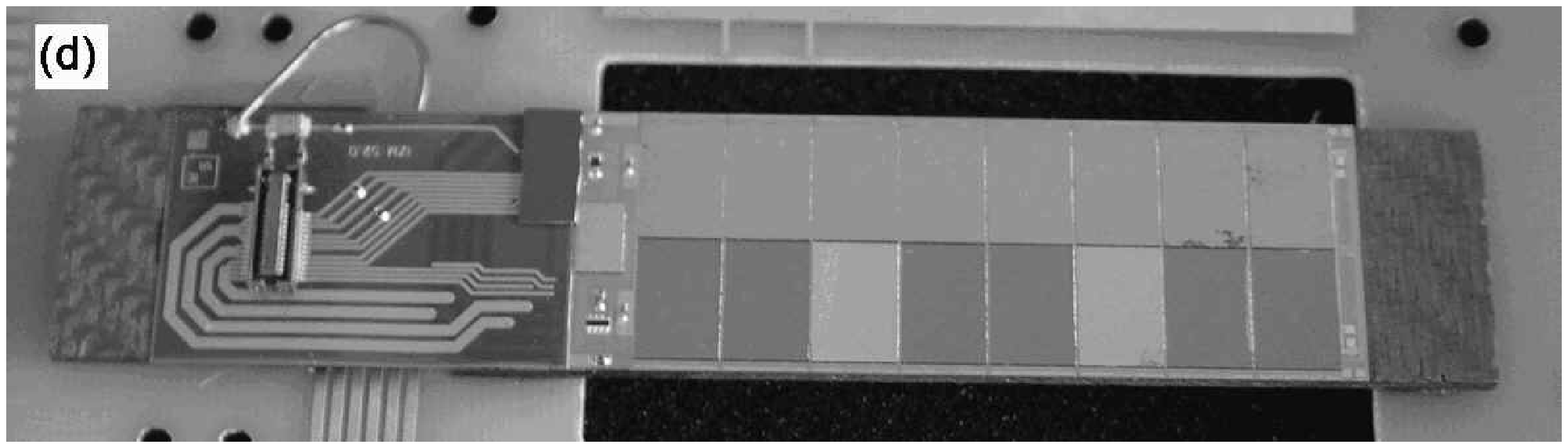}
  \caption{(top left, (a)) Schematic view of a hybrid pixel module and
  (top left, (b)) schematic layout of a MCM-D pixel module indicating the buried via
  structure, (top right) SEM photograph of a MCM-D via structure,
  (bottom) photograph of an ATLAS MCM-D module.
  }
  \label{MCMD}
\end{figure}

The present hybrid-pixel modules of the LHC experiments use an
additional flex-kapton fine-print layer on top of the Si-sensor
(Fig. \ref{MCMD}(a)) to provide power and signal distribution to
and from the module front-end chips. An alternative to the
flex-kapton solution is the so-called Multi-Chip-Module Technology
deposited on Si-substrate (MCM-D) \cite{MCMD1}. A
multi-conductor-layer structure is built up on the silicon sensor.
This allows all bus structures to be buried in four layers in the
inactive area of the module thus avoiding the kapton flex layer
and any wire bonding at the expense of a small thickness increase
of $0.1 \% \ X_0$ (Fig. \ref{MCMD}(b)). The extra freedom in
routing also allows the design of pixel detectors which have the
same pixel dimensions throughout the sensor. Fig. \ref{MCMD}(c)
shows the shows a scanning electron microphotograph (curtesy IZM,
Berlin) of an MCM-D via structure, and Fig. \ref{MCMD}(d) shows
the photograph of an assembled ATLAS MCM-D module \cite{MCMD3}.

\section{Monolithic and Semi-Monolithic Pixel Detectors}
Monolithic pixel detectors, in which amplifying and logic
circuitry as well as the radiation detecting sensor are one
entity, are in the focus of present developments. To reach this
ambitious goal, optimally using a commercially available and cost
effective technology, would be another breakthrough in the field.
So far it has not been completely reached and compromises are
made. The present developments have been much influenced by R$\&$D
for vertex tracking detectors at future colliders such as the
International Linear $e^+e^-$ Collider (ILC) \cite{TESLA-TDR}.
Very low ($\ll$1$\%$ X$_0$) material per detector layer, small
pixel sizes ($\sim$20$\mu$m$\times 20 \mu$m) and a high rate
capability (80 hits/mm$^2$/ms) are required, due to the very
intense beamstrahlung of narrowly focussed electron beams close to
the interaction region, which produce electron positron pairs in
vast numbers. High readout speeds with typical line rates of $50$
MHz and a 40$\mu$s frame time are necessary.

%\begin{figure}[h]
%\begin{center}
%\includegraphics[width=0.22\textwidth]{fig11a.eps}
%%\vskip 0.5cm
%\includegraphics[width=0.22\textwidth]{fig11b.eps}
%\end{center}
%\caption[]{(a) Sketch of a monolithic pixel detector using a high
%resistivity bulk with pMOS transistors at the readout (after
%\cite{parker92}), (b) principle of an Monolithic Active Pixel
%Sensor (MAPS) \cite{MAPS} targeting CMOS electronics with low
%resistivity bulk material. The charge is generated and collected
%by diffusion in the few $\mu$m thin epitaxial Si-layer.}
%\label{fig11}
%\end{figure}

To classify the different monolithic approaches some distinctions
can be made by posing the following questions:
\begin{itemize}
\item does the device allow full CMOS circuitry ? also in the
active sensor area ? \item is the charge collection performed in a
fully depleted bulk providing a large signal ? \item does the
processing technology employ a standard process or are specialized
(non-commercial) technologies necessary ?
\end{itemize}
The above mentioned ultimate monolithic goal would be fulfilled
with a full CMOS commercial standard device with charge collection
in a fully depleted bulk.

The present developments can then be characterized accordingly.

\begin{itemize}
\item \emph{\bf Non-standard CMOS on high resistivity bulk} \\
The first monolithic pixel detector was successfully operated in a
particle beam already in 1992 \cite{parker92} using a high
resistivity p-type bulk p-i-n detector in which the junction had
been created by an n-type diffusion layer. On one side, an array
of ohmic contacts to the substrate served as collection
electrodes. Due to this, only pMOS transistor circuits sitting in
n-wells were possible to be integrated in the active area. The
technology was certainly non-standard and non-commercial. No
further development emerged.
\hfill \\
\item \emph{\bf CMOS technology with charge collection in epi-layer} \\
In some CMOS technologies a lightly doped epitaxial silicon layer
of a few to 15$\mu$m thickness between the low resistivity silicon
bulk and the planar processing layer can be used for charge
collection \cite{meynants98,MAPS1,MAPS2}. The generated charge is
kept in a thin epi-layer atop the low resistivity silicon bulk by
potential wells at the boundary and reaches an n-well collection
diode by thermal diffusion (cf. Fig. \ref{MAPS}(a)). The sensor is
depleted only directly under the n-well diode. The signal charge
is hence very small ($<$1000e) and mostly incomplete; low noise
electronics is the challenge in this development. As a pay-off,
the fabrication of such pixel sensors is potentially very cheap.
With small pixel cells collection times in the order of 100 ns -
150 ns are obtained. Despite using CMOS technology, the potential
of full CMOS circuitry in the active area is not available (only
nMOS) because of the n-well/p-epi collecting diode which does not
permit other n-wells. CMOS monolithic active pixel sensors are
similar to CMOS camera chips, but they are larger in area and must
have a 100$\%$ fill factor for efficient particle detection. The
epi-layer is -- technology dependent -- at most 15$\mu$m thick and
can also be completely absent. Collaborating groups around
IReS$\&$LEPSI \cite{LEPSI1,LEPSI-Portland}, RAL
\cite{RAL-Vertex03}, Irvine-LBNL-Ohio \cite{kleinfelder03}, and
Hawaii-KEK \cite{Varner04} use similar approaches to develop large
scale CMOS active pixels also called MAPS (Monolithic Active Pixel
Sensors) \cite{MAPS1}. Prototype detectors have been produced in
$0.6 \mu$m, $0.35 \mu$m and $0.25 \mu$m CMOS technologies
\cite{Dulinski03,AGay03}.

\begin{figure}[h]
\begin{center}
\includegraphics[width=0.6\textwidth]{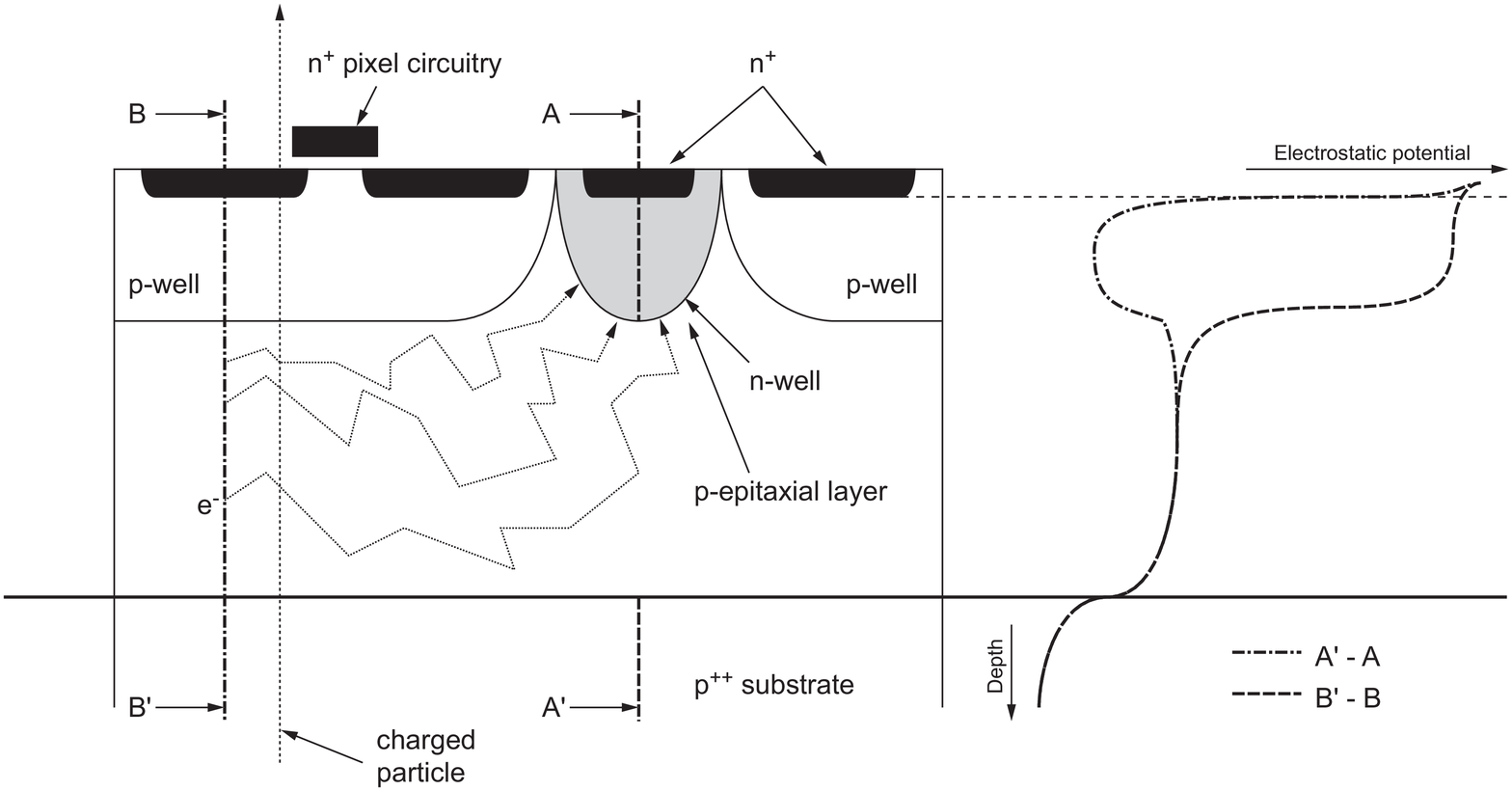}
\hskip 0.5cm
\includegraphics[width=0.35\textwidth]{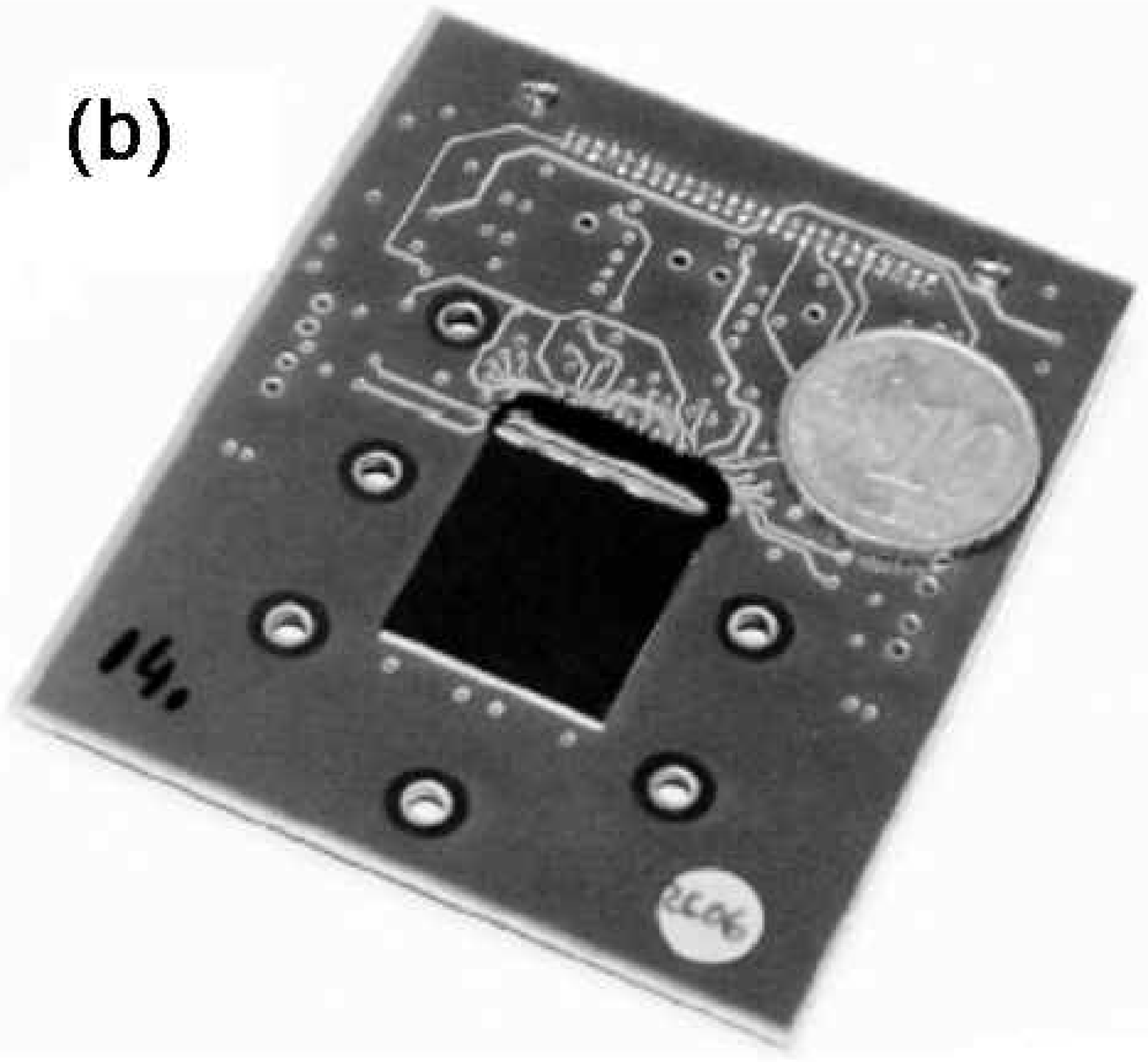}
\end{center}
\caption[]{(a) principle of an Monolithic Active Pixel Sensor
(MAPS) \cite{MAPS1} targeting CMOS electronics with low
resistivity bulk material. The charge is generated and collected
by diffusion in the very few $\mu$m thick epitaxial Si-layer. (b)
MAPS detector with 100 nm thin backside entrance window.}
\label{MAPS}
\end{figure}

Matrix readout of MAPS is performed using a standard 3-transistor
circuit (line select, source-follower stage, reset) commonly
employed by CMOS matrix devices, but can also include current
amplification and current memory \cite{Dulinski03}. For an image
two complete frames are subtracted from each other (CDS) which
suppresses switching noise. Noise figures of 10-30e and S/N
$\sim$20 have been achieved with spatial resolutions below
5$\mu$m. Regarding radiation hardness MAPS appear to sustain
non-ionizing radiation (NIEL) to $\sim$10$^{12}$n$_{eq}$ while the
effects of ionizing radiation damage (IEL) are at present still
under investigation. The present focus of further development lies
in making larger area devices for instance by stitching over
reticle boundaries \cite{AGay03}, increasing the charge collection
performance in the epi-layer by triple-well \cite{RAL-Vertex03},
photo-gate \cite{Kleinfelder-Portland}, and photo-FET
\cite{Dulinski03,LEPSI-Portland} techniques and developing a
higher radiation tolerance. In addition, for applications like
{\it e.g.} precise beam position monitoring in hadron therapy,
devices with very thin entrance windows are needed to detect
$\sim$20 keV electrons scattering off a thin metal foil held in a
hadron beam. Such a thinned MAPS detector, which is also capable
of autoradiographic tritium detection, is shown in Fig.
\ref{MAPS}(b).

Above all, the advantages of a fully CMOS monolithic device relate
to the adoption of standard VLSI technology and its resulting low
cost potential ($\sim$25$\$$ per cm$^2$). In turn the
disadvantages are also largely related to the dependence on
commercial standards. The thickness of the epi-layer varies for
different technologies. It is thinner for processes with smaller
structure sizes and along with this also the signal decreases.
Only a few processing technologies are suited. With the rapid
change of commercial process technologies this is an issue of
concern. Furthermore, as already stated, in the active area, due
to the n-well collection diode, only nMOS circuitry is possible.
The voltage signals are very small ($\sim$mV), of the same order
as transistor threshold dispersions requiring very low noise VLSI
design. The radiation tolerance of CMOS pixel detectors for
particle detection beyond 10$^{12}$ n$_{eq}$ is still a problem
although the achieved tolerance should be well sufficient for use
in a future Linear Collider environment. Improved readout concepts
and device development for high rate particle detection at a
linear collider are under development \cite{TESLA-TDR}. Concrete
project planning to use CMOS active pixels in real experiments are
the upgrades to the STAR micro vertex detector \cite{Dulinski03}
and to the Super Belle detector (innermost layer) at KEK, for
which very encouraging prototype developments and results have
been presented at this conference \cite{Varner04}.

\hfill \\
\item \emph{\bf CMOS on SOI (non-standard)} \\
In order to exploit high resistivity bulk material, the authors of
\cite{SOI-Portland} develop pixel sensors with full CMOS circuitry
using the Silicon-on-Insulator (SOI) technology. The connection to
the charge collecting bulk is done by vias (Fig. \ref{SOI}(left)).
The technology offers full charge collection in 200--300$\mu$m
high resistive silicon with full CMOS electronics on top. At
present the technology is however definitely not a commercial
standard and the development is still in its beginnings. More
details on this interesting new technology has been reported at
this conference and can be found in \cite{SOI_Hiroshima}.

\begin{figure}[h]
\begin{center}
\includegraphics[width=0.45\textwidth]{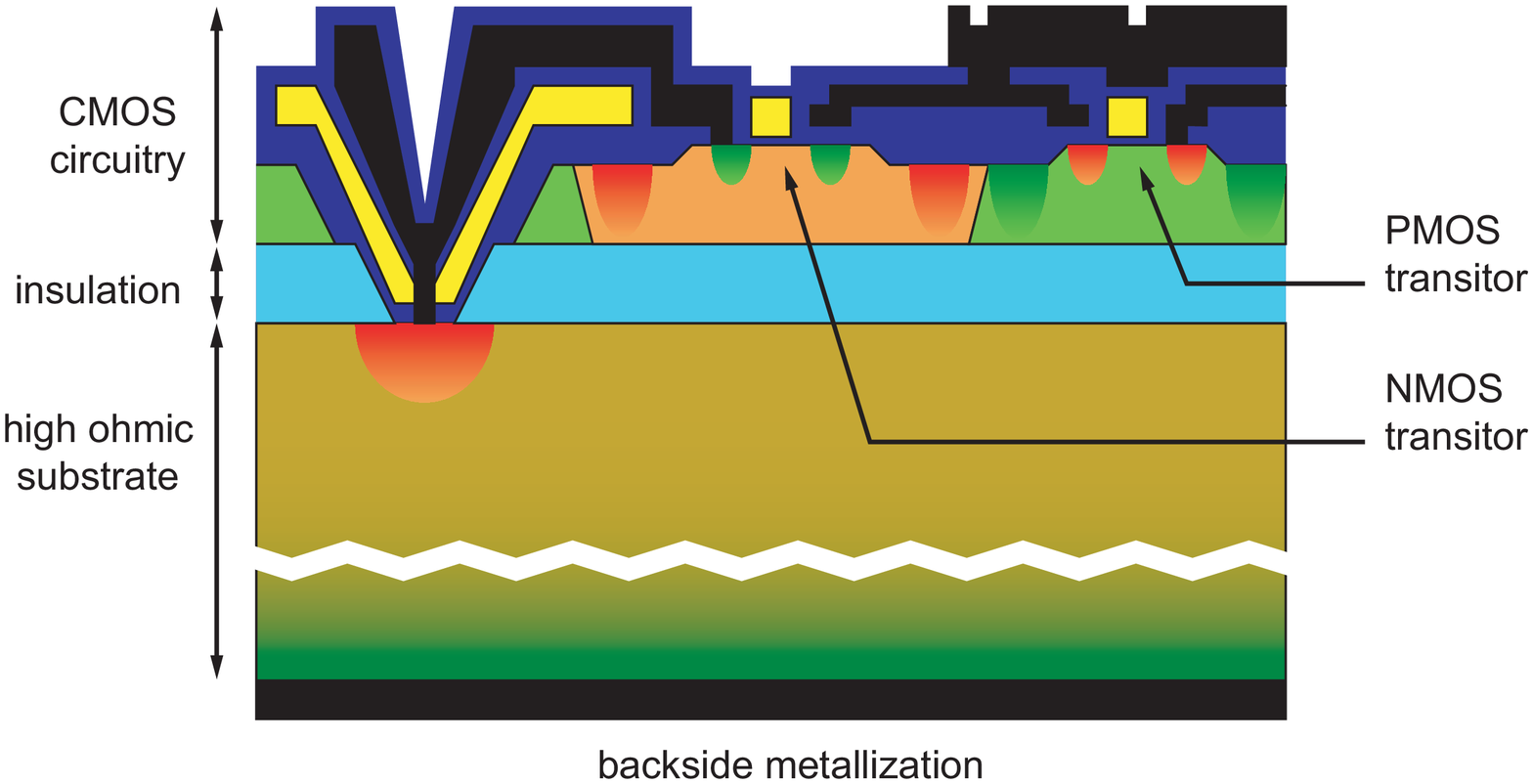}
\hskip 0.5cm
\includegraphics[width=0.40\textwidth]{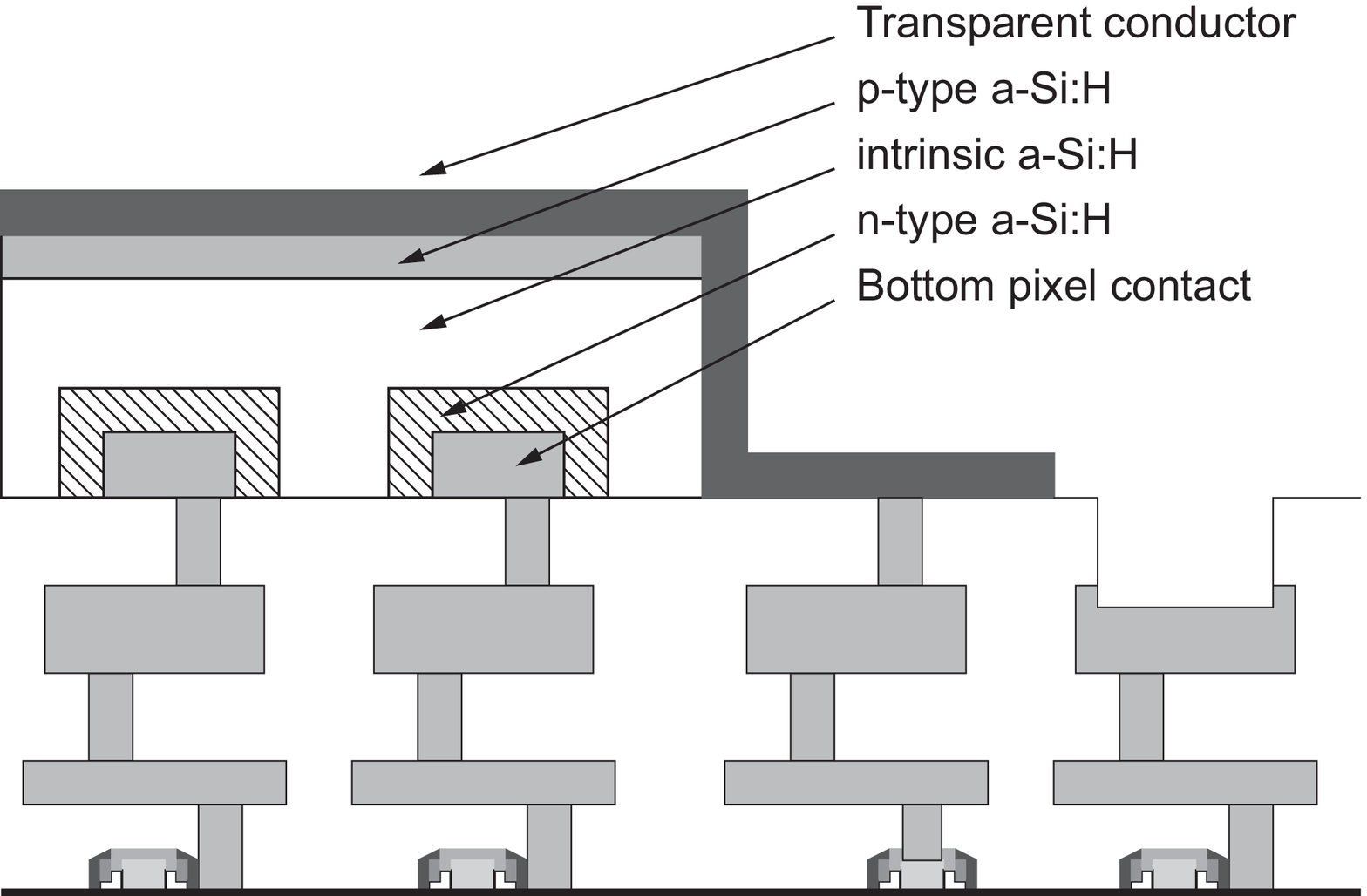}
\end{center}
\caption[]{(left) Cross section through a monolithic CMOS on SOI
pixel detector using high resistivity silicon bulk insolated from
the low resistivity CMOS layer with connecting vias in between
\cite{HAPS-SOI,SOI-Portland}, (right) cross section through a
structure using amorphous silicon on top of standard CMOS VLSI
electronics \cite{theil2001,jarron02}.} \label{SOI}
\end{figure}

\hfill \\
\item \emph{\bf Amorphous silicon on standard CMOS ASICs} \\
Hydrogenated amorphous-Silicon (a-Si:H), where the H-content is up
to 20$\%$, can be put as a film on top of CMOS ASIC electronics.
a-Si:H has been studied as a sensor material long ago and has
gained interest again \cite{theil2001,jarron02} with the
advancement in low noise, low power electronics. The signal charge
collected in the $<$30$\mu$m thick film is in the range of
500-1500 electrons. A cross section through a typical a-Si:H
device is shown in Fig. \ref{SOI}(right). From a puristic view it
is more a hybrid technology, but the main disadvantage, the
hybridization connection, is absent. The radiation hardness of
these detectors appears to be very high $>$10$^{15}$cm$^{-2}$ due
to the defect tolerance and defect reversing ability of the
amorphous structure and the larger band gap (1.8 eV). The carrier
mobility is very low ($\mu_e$ = 2-5 cm$^2$/Vs, $\mu_h$ = 0.005
cm$^2$/Vs), i.e. essentially only electrons contribute to the
signal. Basically any CMOS circuit and IC technology can be used
and technology changes are not critical for this development. For
high-Z applications poly-crystalline HgI$_2$ constitutes a
possible semiconductor film material. The potential advantages are
small thickness, radiation hardness, and low cost. The development
is still in its beginnings and -- as for CMOS active pixel sensors
-- a real challenge to analog VLSI design.

\hfill \\
\item \emph{\bf Amplification transistor implanted in high resistivity bulk} \\
In so-called DEPFET pixel sensors \cite{kemmer87} a JFET or MOSFET
transistor is implanted in every pixel on a sidewards depleted
\cite{gatti84} bulk. Electrons generated by radiation in the bulk
are collected in a potential minimum underneath the transistor
channel thus modulating its current (Fig. \ref{DEPFET_principle}).
The bulk is fully depleted rendering large signals. The small
capacitance of the internal gate offers low noise operation. Both
together can be used to fabricate thin devices. The sensor
technology is non-standard and the operation of DEPFET pixel
detectors requires separate steering and amplification ICs.

\begin{figure}[h]
\begin{center}
\includegraphics[width=0.8\textwidth]{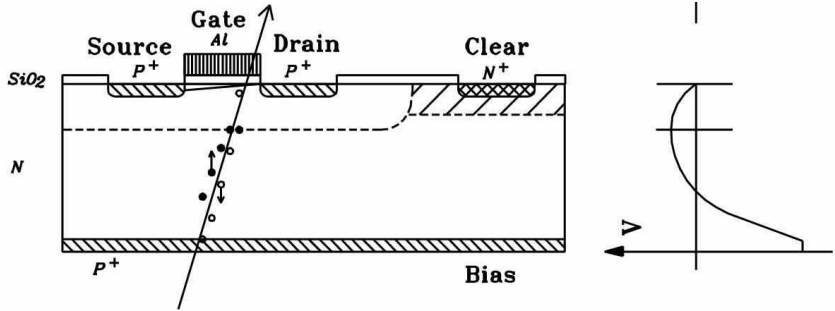}
%\vskip 0.5cm
%\includegraphics[width=0.18\textwidth]{fig12b.eps}
\end{center}
\caption[]{Principle of operation of a DEPFET pixel structure
based on a sidewards depleted detector substrate material with an
imbedded planar field effect transistor. Cross section (left) of
half a pixel with symmetry axis at the left side, and potential
profile (right).} \label{DEPFET_principle}
\end{figure}

The DEPFET detector principle is shown in Fig.
\ref{DEPFET_principle}. Sidewards depletion \cite{gatti84}
provides a parabolic potential which has -- by appropriate biasing
and a so-called deep-n implantation -- a local minimum for
electrons ($\sim 1 \mu$m) underneath the transistor channel. The
channel current can be steered and modulated by the voltage at the
external gate and - important for the detector operation - also by
the deep-n potential (internal gate). Electrons collected in the
internal gate are removed by a clear pulse applied to a dedicated
contact outside the transistor region. The very low input
capacitance ($\sim$ few fF) and the in situ amplification makes
DEPFET pixel detectors very attractive for low noise operation
\cite{ulrici03}. Amplification values of 400 pA per electron
collected in the internal gate have been achieved. Further current
amplification and storage enters at the second level stage.

DEPFET pixels are currently being developed for three very
different application areas: vertex detection in particle physics
\cite{DEPFET-TESLA,DEPFET_Portland}, X-ray astronomy \cite{holl02}
and for biomedical autoradiography \cite{ulrici03}. With round
single pixel structures noise figures of 2.2e at room temperature
and energy resolutions of 131 eV for 6 keV X-rays have been
obtained \cite{DEPFET_Portland}. With small (20x30 $\mu$m$^2$)
linear structures fabricated for particle detection at a Linear
Collider (ILC) the noise figures are about 10e. The spatial
resolution in matrices operated with 50 kHz line rates is
\cite{DEPFET_Portland,ulrici03} $\sigma = (4.3 \pm 0.8) \mu$m (57
LP/mm) for 22 keV $\gamma$ and $\sigma = (6.7 \pm 0.7) \mu$m (37
LP/mm) for 6 keV $\gamma$. The capability to observe tritium in
autoradiography applications has been demonstrated
\cite{ulrici03,DEPFET_Portland}.

\begin{figure}[htb]
\begin{center}
\includegraphics[width=1.0\textwidth]{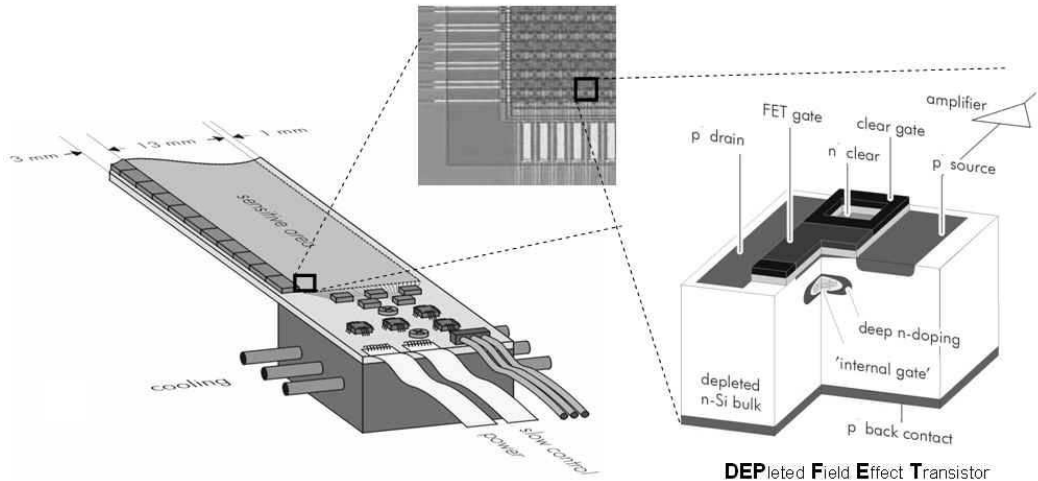}
\vskip 0.5cm
\includegraphics[width=0.7\textwidth]{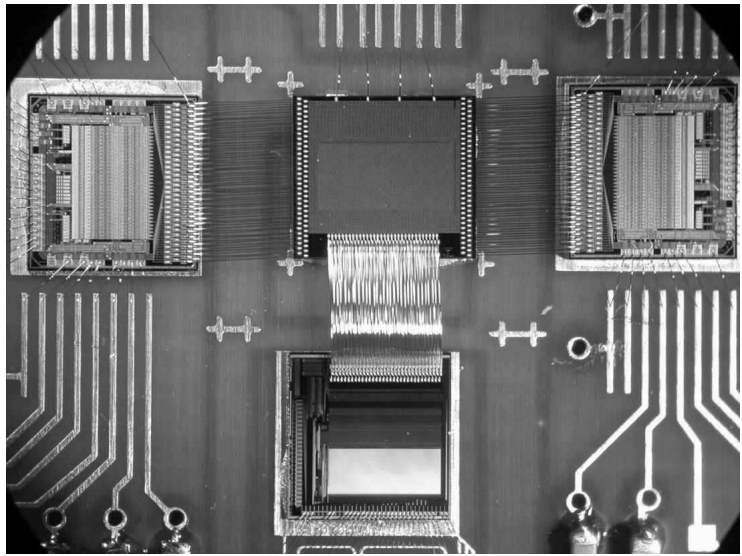}
\end{center}
\caption[]{(top) Sketch of a ILC first layer module with thinned
sensitive area supported by a silicon frame. The enlarged view
show a DEPFET matrix and a DEPFET double pixel structure,
respectively, (bottom) photo of a DEPFET matrix readout system for
LC applications. The sequencer chips (SWITCHER II) for select and
clear are placed on the sides of the matrix, the current readout
chip (CURO II) at the bottom.} \label{DEPFET_TESLA}
\end{figure}

The very good noise capabilities of DEPFET pixels are very
important for low energy X-ray astronomy and for autoradiography
applications. For particle physics, where the signal charge is
large in comparison, this feature is used to design very thin
detectors ($\sim$50$\mu$m) with very low power consumption when
operated as a row-wise selected matrix
\cite{Trimpl02,DEPFET_Portland}. Thinning of sensors to a
thickness of 50$\mu$m using a technology based on deep anisotropic
etching has been successfully demonstrated \cite{laci04}. For the
development of DEPFET pixels for a Linear Collider, matrix row
rates of 50 MHz and frame rates for 520x4000 pixels of 25 kHz,
read out at two sides, are targeted. Sensors with cell sizes of
20x30$\mu$m$^2$ have been fabricated and complete clearing of the
internal gate, which is important for high speed on-chip pedestal
subtraction, has been demonstrated \cite{DEPFET_Portland}. A
sketch of a first layer module made of DEPFET sensors is shown in
fig. \ref{DEPFET_TESLA}(top). A large matrix is readout using
sequencer chips for row selection and clear, and a column readout
chip based on current amplification and storage
\cite{Trimpl02,DEPFET_Portland}. Both chips have been developed at
close to the desired speed for a Linear Collider. Figure
\ref{DEPFET_TESLA}(bottom) shows a DEPFET pixel matrix readout
system suited for readout speeds at the ILC. The estimated power
consumption for a five layer DEPFET pixel vertex detector at the
ILC is only 5W. This assumes a power duty cycle of 1:200 and no
power consumption of the SWITCHER chip in the off state. Such a
performance renders a very low mass detector without cooling pipes
feasible. The most recent structures use MOSFETs as DEPFET
transistors for which the radiation tolerance still has to be
investigated.
\end{itemize}

%\begin{figure}[htb]
%\begin{center}
%\includegraphics[width=1.0\textwidth]{figs/DEPFET_system.eps}
%\end{center}
%\caption[]{Photo of a DEPFET matrix readout system for LC
%applications. The sequencer chips (SWITCHER II) for select and
%clear are placed on the sides of the matrix, the current readout
%chip (CURO II) at the bottom.} \label{DEPFET_system}
%\end{figure}
%
\section{Summary}
The Hybrid Pixel technology, in which sensor and electronic chip
are separate entities connected via bump bonding techniques, has
shown the way to large area ($\sim$m$^2$) pixel detectors for
experiments at the LHC. These detectors are in construction and
the maturity of the technology, including radiation tolerance to
500 kGy doses, has been proven. New developments include
interleaved pixels having different pixel- and readout pitches,
MCM-D structures, and the use of diamond as a sensor. Hybrid
pixels in which radiation quanta are individually counted and
summed over an exposure time have opened a new approach in
radiology imaging, especially at low dose rates. Monolithic or
semi-monolithic detectors, in which detector and readout
ultimately are one entity, are currently being developed in
various forms, largely driven by the needs for particle detection
at future colliders. CMOS active pixel sensors using standard
commercial technologies on low resistivity bulk and SOI pixels,
a-SI:H, and DEPFET-pixels, which try to maintain high bulk
resistivity for charge collection, constitute the different
efforts to monolithic detectors which are presently being carried
out.

\subsubsection*{Acknowledgements}
The author would like to thank W.~Kuzewicz, M.~Caccia,
W.~Dulinski, M. Keil, R.~Kohrs, M.~Trimpl, E. Radermacher,
R.~Richter, T.~Rohe, and G.~Stefanini for providing new material
and information for this review. He is also very grateful to
G.~Varner for a scrutinizing reading of the manuskript.

\bibliographystyle{unsrt}
\bibliography{wermes_Hiroshima2004}

\end{document}